\documentclass[a4paper,11pt]{article}
\usepackage{jheppub}

\pdfoutput=1

\usepackage{graphicx} 
\usepackage{dcolumn}  
\usepackage{bm}       

\usepackage{amsmath}
\usepackage{amssymb}
\usepackage{booktabs}
\usepackage[caption=false]{subfig}
\usepackage[
colorlinks=true,
citecolor=blue
]{hyperref}
\usepackage[dvipsnames]{xcolor}
\usepackage{multirow}
\usepackage{enumitem} 

\usepackage{svg}

\graphicspath{{img/}} 

\setlist[description]{font=\normalfont\space}

\newcommand{\zjet}{\ensuremath{Z+}\text{jet }}

\newcommand{\redkappa}[2]{\ensuremath{
\kappa(\text{\footnotesize #1}|\text{\footnotesize #2})
}}

\begin{document}

\title{Quark-versus-gluon tagging in CMS Open Data with CWoLa and TopicFlow}

\date{\today}

\author[a]{Matthew J.\ Dolan}
\emailAdd{dolan@unimelb.edu.au}
\affiliation[a]{ARC Centre of Excellence for Dark Matter Particle Physics, School of Physics, The University of Melbourne, Victoria 3010, Australia}

\author[b]{John Gargalionis}
\emailAdd{john.gargalionis@ific.uv.es}
\affiliation[b]{Departamento de F\'isica Te\'orica and IFIC, Universidad de Valencia-CSIC, C/Catedr\'atico Jos\'e Beltr\'an, 2, E-46980 Paterna, Spain}

\author[c]{Ayodele Ore}
\emailAdd{ore@thphys.uni-heidelberg.de}
\affiliation[c]{Institut f\"ur Theoretische Physik, Universit\"at Heidelberg, Germany}

\abstract{We use the CMS Open Data to examine the performance of weakly-supervised learning for tagging quark and gluon jets at the LHC. We target $Z$+jet and dijet events as respective quark- and gluon-enriched mixtures and derive samples both from data taken in 2011 at 7 TeV, and from Monte Carlo. CWoLa and TopicFlow models are trained on real data and compared to fully-supervised classifiers trained on simulation. In order to obtain estimates for the discrimination power in real data, we consider three different estimates of the quark/gluon mixture fractions in the data. Compared to when the models are evaluated on simulation, we find reversed rankings for the fully- and weakly-supervised approaches. Further, these rankings based on data are robust to the estimate of the mixture fraction in the test set. Finally, we use TopicFlow to smooth statistical fluctuations in the small testing set, and to provide uncertainty on the performance in real data.}

\maketitle
\flushbottom
\section*{Introduction}

Discriminating between the quark or gluon origins of hadronic jets is a classic and difficult problem in the phenomenology of quantum chromodynamics (QCD) at particle colliders, which has been extensively studied experimentally~\cite{DELPHI:1999gah,OPAL:1999jkz,CDF:2005prv,ATLAS:2016wzt,ATLAS:2017nma,ATLAS:2017dfg,CMS:2017yer,CMS:2021iwu,ATLAS:2023pdx}. There are multiple reasons it would be desirable to do this. From the perspective of Beyond the Standard Model physics, tagging hadronic decays of potential resonances could be helpful in reconstructing the underlying UV theory. From a Standard Model (SM) perspective, studies that aim to isolate Higgs production from vector boson fusion, where a Higgs is produced in association with two quark jets, must determine the contamination from gluon fusion, which can involve Higgs production with multiple gluon jets.

Like other jet classification tasks, there has been extensive study of neural-network-based classifiers applied to  quark/gluon discrimination~\cite{Komiske:2016rsd,ATLAS:2017dfg,Cheng:2017rdo,Luo:2017ncs,Kasieczka:2018lwf,Komiske:2018cqr,Lee:2019cad,Lee:2019ssx,Moreno:2019bmu,Qu:2019gqs,Mikuni:2020wpr,Dreyer:2020brq,Bogatskiy:2022czk,Butter:2022xyj,Qu:2022mxj,He:2023cfc}. These models, built on modern deep-learning architectures, are able to achieve discrimination power far exceeding traditional approaches based on jet substructure observables. However, typically they are trained and tested in a fully-supervised setting, making use of parton-level information in Monte-Carlo (MC). Given that these taggers owe their strength to a sensitivity to the detailed correlations in the substructure of a jet, they are at risk of overfitting to mismodelled features in Monte Carlo (MC) samples. In particular, quark/gluon discrimination is known to be highly sensitive to non-perturbative effects in QCD~\cite{Gras:2017jty, Dreyer:2021hhr}. Combining this with the fact that contemporary simulations for parton showering and hadronisation have large theoretical uncertainties~\cite{mo:2017gzp}, we are motivated to explore methods for training and evaluating quark/gluon taggers on real data.

Such methods can be realised within the paradigm of weakly-supervised learning, which broadly covers methods that take as input datasets with noisy labels or, equivalently, \emph{mixed} classes. Such datasets are generally easier to obtain than the completely separated samples required for full supervision. Indeed, with motivated phase-space selections, one can isolate regions enriched with quark or gluon jets~\cite{Gallicchio:2011xc} at the parton level. These selections can then be applied in data, yielding input samples for weakly-supervised algorithms. One such method is Classification Without Labels (CWoLa, pronounced ``koala'')~\cite{Metodiev:2017vrx, Komiske:2018oaa}, where a classifier is tasked with discriminating two datasets of different mixture proportions. Under certain assumptions, this kind of classifier can be expected to match the performance of a fully-supervised model. In contrast to previous weakly-supervised paradigms~\cite{Dery:2017fap}, this is possible even without knowledge of the class proportions. Previous works employing CWoLa for quark/gluon classification include~Refs.~\cite{Lee:2019ssx,ATLAS:2020iwa,Komiske:2022vxg,CMS:2024nsz}.\footnote{See also Ref.~\cite{Alvarez:2021zje} for an unsupervised approach.}

More recent work on \emph{jet topics}~\cite{Metodiev:2018ftz,Komiske:2018vkc,Komiske:2022vxg} offers a framework within which the mixture fractions can be estimated in data, leading to predictions for the individual quark- and gluon-like distributions. This technique provides an ‘operational’ definition of the terms ‘quark jet’ and ‘gluon jet’ and has seen application in both experimental~\cite{ATLAS:2019rqw, Brewer:2020och, Ying:2022jvy, LeBlanc:2022bwd} and theoretical~\cite{Larkoski:2019nwj, Stewart:2022ari,Takacs:2021bpv} settings. In a previous work~\cite{Dolan:2022ikg} some of the authors introduced a deep generative approach to jet topics, called TopicFlow. There, we showed that normalising flows can demix quark and gluon distributions in many dimensions and demonstrated their advantage by smoothing statistics with oversampling. While TopicFlow is compatible with use
on real data, previously it was only tested on simulated collision events at the Large Hadron Collider (LHC).

The CMS collaboration has made a wide range of data collected by the LHC publicly available through the CERN Open Data Portal~\cite{cernopendataportal}. This now includes fully-reconstructed collision data as well as detector-simulated MC samples, paving the way for exploratory analysis strategies, measurements, and new data formats and frameworks to be investigated outside of the experimental collaborations. Work in these directions has already begun, with jet studies~\cite{Larkoski:2017bvj, Tripathee:2017ybi,Komiske:2022vxg,Komiske:2022enw,Arias:2023wai} as well as machine-learning studies~\cite{Madrazo:2017qgh,Andrews:2018nwy, Andrews:2019faz, knapp2020adversarially} and new search strategies and techniques~\cite{PaktinatMehdiabadi:2019ujl, Cesarotti:2019nax, Komiske:2019jim,Lester:2019bso, Apyan:2019ybx, An:2021yqd}.

In this paper, we make use of the CMS Open Data to extend investigations of weakly-supervised learning using TopicFlow.  A previous work~\cite{Komiske:2022vxg} has performed a similar study, defining quark- and gluon-enriched samples by partitioning QCD jets in pseudorapidity (see also ~\cite{ATLAS:2023pdx}). Another possibility, recently advocated in~\cite{Baron:2023hkp}, is to combine measurements made at different energies. In contrast to those works, and similarly to CMS, we use distinct event topologies for our sources of enriched samples. Specifically, we isolate \zjet and dijet samples which are respectively expected to be enriched in quark jets and gluon jets. Using both simulated and real events, we extract jet topics and train CWoLa and TopicFlow models. Significance improvement curves evaluated on MC are compared with estimates for the curves in data, revealing reversed rankings for the fully- and weakly-supervised approaches.

The remainder of the paper is organised as follows. In Section~\ref{sec:datasets} we detail the selections that define our enriched samples in simulation and collision data. Section~\ref{sec:weak} describes the weakly-supervised methods that we apply to the datasets, with specific implementations outlined in Section~\ref{sec:training}. We present the results of our study in Section~\ref{sec:results} before providing concluding remarks in Section~\ref{sec:conclusion}.

\section{Datasets and Event Selection}
\label{sec:datasets}

In the following we discuss our datasets and event selection. A discussion of the CMS software framework, our data-analysis pipeline and some background on data taking at CMS are presented in Appendix~\ref{sec:TheCMSOpenData}. The appendix can be skipped by the reader uninterested in procedural details related to the CMS Open Data.

The experimental data we use for our analysis was collected by the CMS experiment in 2011 and made publicly available through the CERN Open Data Portal~\cite{cernopendataportal} in 2016. It includes $2.5 \text{ fb}^{-1}$ of $7$~TeV proton--proton collisions collected in Run 2011A, as well as simulated events which have been passed through the CMS detector simulation. This was the second of a number of batches of data published regularly by the collaboration since 2014. Currently, more than 2~PB from the 2010--2012 run period have been made publicly available, including all data collected in 2010 and 2011.

Our study makes use of four datasets: quark-enriched and gluon-enriched samples from both collision data and Monte Carlo (MC) simulation. For the gluon-enriched datasets, we use the {\tt Jet} primary dataset from RunA of 2011~\cite{jet2011A} and the SM-exclusive QCD simulation generated with $Z2$-tuned {\sc Pythia 6}~\cite{Sjostrand:2006za}. These simulated data are accessed through separate samples organised by $p_T$ ranges~\cite{qcdsim15to30, qcdsim30to50, qcdsim50to80, qcdsim80to120, qcdsim120to170, qcdsim170to300, qcdsim300to470, qcdsim470to600, qcdsim600to800, qcdsim800to1000, qcdsim1000to1400, qcdsim1400to1800, qcdsim1800}. All of the jet triggers in the MC fire with unit prescales. A detailed discussion of the CMS Run 1 Trigger system can be found in~\cite{CMS:2016ngn}. We consider jets with $p_{T} > 30$~GeV and $|\eta|<2.1$ that satisfy medium jet quality criteria (JQC)~\cite{CMS:2011shu}. Events with fewer than two such jets are discarded and we also require the {\tt DiJetAve30} high-level trigger to fire in the collision events. For the quark-enriched datasets, we use the {\tt DoubleMu} primary dataset from RunA and RunB of 2011~\cite{doublemu2011A} and the SM-inclusive Drell--Yan (DY) MC, also generated with $Z2$-tuned {\sc Pythia 6}~\cite{dymc}. We take events that contain two global tracker muons~\cite{CMS:2018rym} satisfying $p_{T,\mu}> 26$~GeV and $|\eta_\mu| < 2.4$. The leading dimuon system is identified as a $Z$ candidate and required to have 70~GeV $\leq m_{\mu\mu} \leq$ 110~GeV. After discarding jets that overlap with either muon within their radii, we use the same jet definition and cuts as for the gluon-enriched sample. Events are further subjected to $\Delta \phi(j_0, \mu\mu) > 2$, ensuring that the selected events are dominantly two-body processes ($Z+j$) with minimal additional jet activity. We also apply a $p_T$ asymmetry cut $\frac{|p_{T,\mu\mu} - p_{T, j_0}|}{p_{T,\mu\mu} + p_{T, j_0}} < 0.3$ where $j_0$ refers to the leading jet~\cite{CMS:2021iwu}. This limits contributions from large QCD corrections associated with events where $p_{j1,T}\gg p_{Z,T}$ ~\cite{Rubin:2010xp}. For the {\tt DoubleMu} dataset, we require a firing \texttt{Mu13\_Mu8} high-level trigger, which is not prescaled. In all datasets, we select events where the leading jet has $p_T$ in the range $[50,80]$~GeV. In the two simulated datasets, we remove jets that are not assigned a parton match, which is the case for approximately 15\% of the jets in QCD MC and 3\% in DY MC~(see Appendix~\ref{app:parton-matching} for further discussion). Finally, the parton labels are used to construct `pure' quark and gluon datasets from the QCD simulation which will facilitate supervised training. The number of events passing selection in each dataset is given in Table~\ref{tab:dataset-sizes}.

\begin{table}
    \centering
    \begin{tabular}{rrrr}
        \toprule
        \textbf{Dataset}  & \textbf{Total events} & \textbf{Quarks} & \textbf{Gluons}\\\midrule
        \texttt{DoubleMu} & 41,773 & --- & ---                   \\
        \texttt{Jet}      & 82,162 & --- & ---                   \\
        DY MC      & 95,324 & 70,568 & 24,756 \\
        QCD MC            & 3,064,713 & 868,556 & 2,196,157 \\         
        \bottomrule
    \end{tabular}
    \caption{Summary of the number of events remaining in each dataset as well as the flavour of the leading jet (MC only) after all selections described in Section~\ref{sec:datasets}.}
    \label{tab:dataset-sizes}
\end{table}

\begin{figure}
    \centering
    \subfloat[]{
        \includegraphics[width=0.45\textwidth]{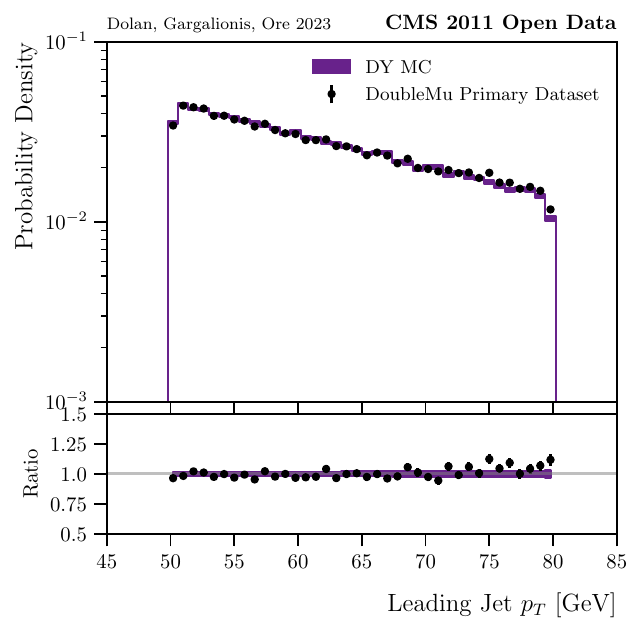}
    }\hfill
    \subfloat[]{
        \includegraphics[width=0.45\textwidth]{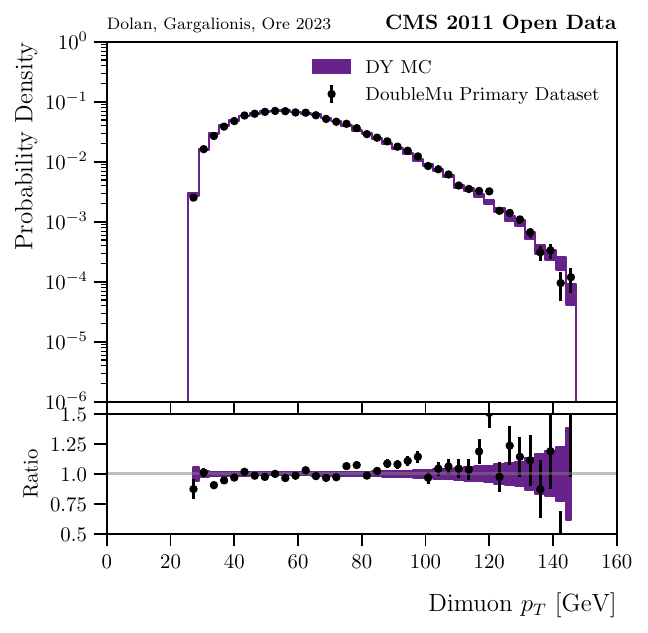}
    }\\
    \subfloat[]{
        \includegraphics[width=0.45\textwidth]{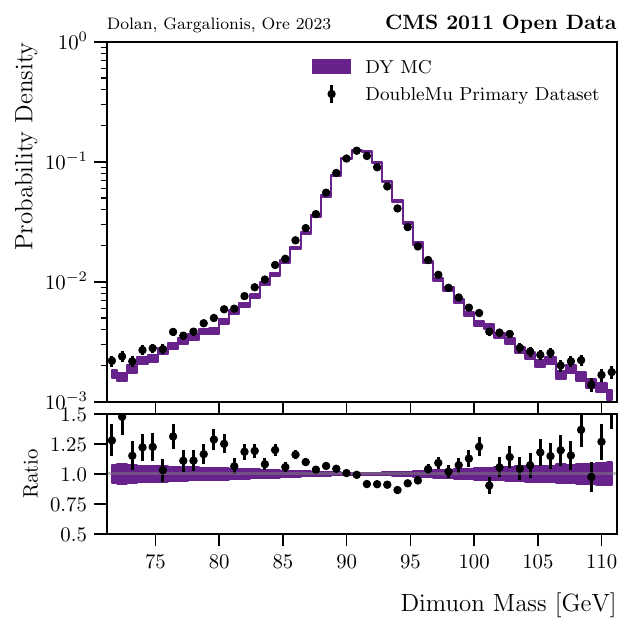}
    }\hfill
    \subfloat[]{
        \includegraphics[width=0.45\textwidth]{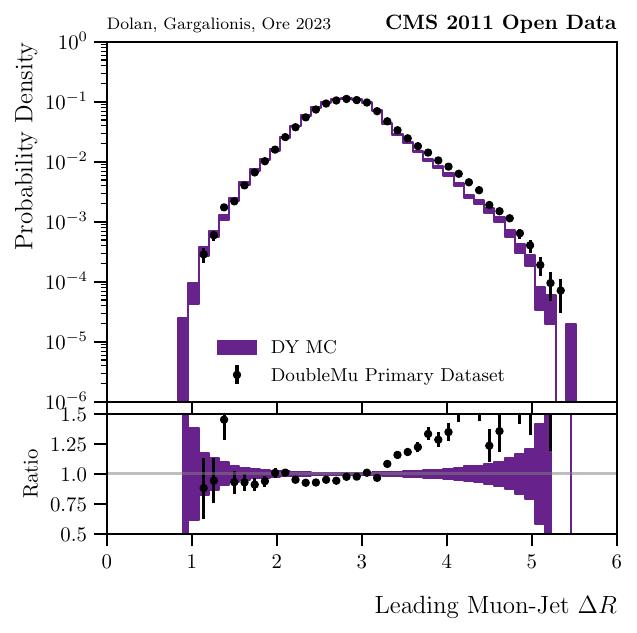}
    }
    \caption{Validation of the selection cuts applied to the \zjet samples. The panels show a) leading jet $p_T$, b) dimuon $p_T$, c) dimuon invariant mass and d) leading muon-jet separation $\Delta R$. The DY Monte Carlo (from Pythia) is in purple, the data is the black points. The error bars show statistical uncertainties. All distributions are normalised to unit area.}
    \label{fig:zjet-validation}
\end{figure}
In Figure~\ref{fig:zjet-validation} we show validation plots for our event selection and workflow for the $Z+$jet samples. We compare the DY Monte Carlo (in purple) with the data from the \texttt{DoubleMu} Primary dataset (black data-points). All distributions have been normalised by area. Clockwise from top-left we show the leading jet $p_T$, the dimuon $p_T$, the separation $\Delta R$ between the leading muon and jet, and the dimuon invariant mass. Here, and throughout this paper, uncertainty bands are statistics-only. We find excellent agreement in the leading jet $p_T$ and dimuon $p_T$, although statistics start to become an issue at higher muon $p_T$ values. The dimuon invariant mass is peaked at $m_Z$, although the shape is not perfectly modelled. We also observe some small discrepancies at large muon-jet separation.

\begin{figure}
    \centering
    \subfloat[\label{fig:dijet_jetPt.pdf}]{
        \includegraphics[width=0.45\textwidth]{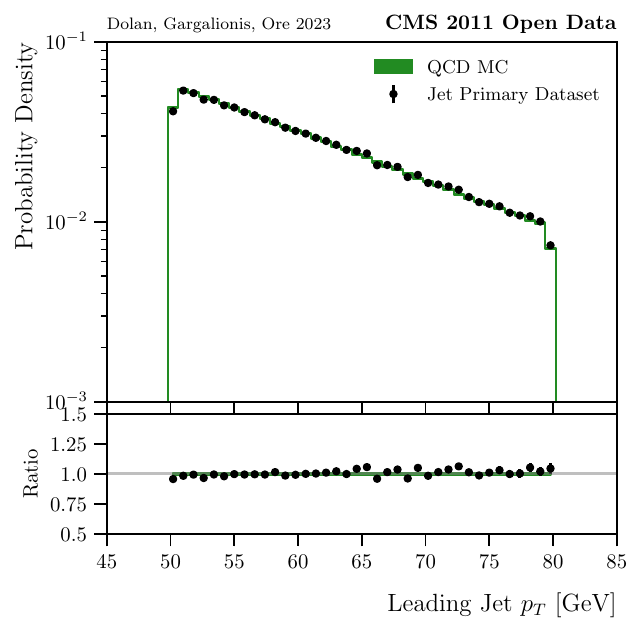}
    }\hfill
    \subfloat[\label{fig:dijet_aveDijetPt.pdf}]{
        \includegraphics[width=0.45\textwidth]{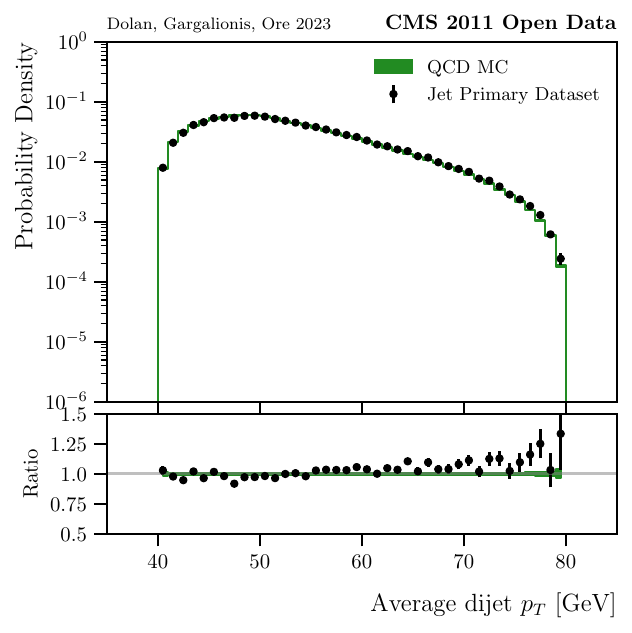}
    }
    \caption{Validation of the selection cuts applied to the dijet samples for \protect\subref{fig:dijet_jetPt.pdf} the leading jet $p_T$ and \protect\subref{fig:dijet_aveDijetPt.pdf} the average dijet $p_T$. The QCD Monte Carlo (from Pythia) is in green and the data is the black points. The error bars show statistical uncertainties. All distributions are normalised by area.}
    \label{fig:dijet-validation}
\end{figure}
Figure~\ref{fig:dijet-validation} shows two distributions derived from the QCD Monte Carlo and the \texttt{Jet} Primary Dataset. The left-hand plot shows the leading jet $p_T$ and right-hand one the average dijet $p_T$. Again, we observe good agreement between the simulation and data.

For our study, we also consider jet substructure observables used for quark/gluon discrimination in previous CMS works~\cite{CMS-PAS-JME-13-002,CMS-PAS-JME-16-003}. We consider the multiplicity of the jet constituents, the jet mass, the jet fragmentation distribution $p_T^D$ and the major axis of the jet $\sigma_1$. Definitions for the latter two observables can be found in Appendix~\ref{app:substructure-definitions}. In Figure~\ref{fig:additional-substructure}, we compare the data and MC distributions for the leading jet substructure. The left column shows dijet events, which exhibit discrepancies on the order of 25\%, notably for multiplicity and $p_T^D$. The agreement is somewhat better for \zjet events in the right column, though systematic deviations remain. Given that we only have access to statistical uncertainties in MC, we cannot determine whether the observed biases are covered by varying systematics such as the parton shower or renormalisation/factorisation scales. We checked that these substructure variables are not correlated to the event-level observables that also displayed biases in Figure~\ref{fig:zjet-validation}.
\begin{figure}
    \subfloat[\label{fig:dijet_nConstits}]{
            \includegraphics[width=0.45\textwidth]{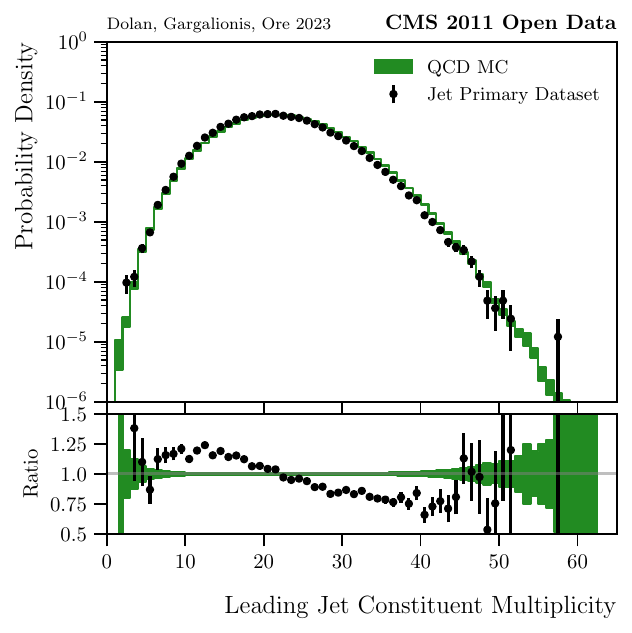}
        }
    \hspace{1cm}
    \subfloat[\label{fig:zjet_nConstits}]{
        \includegraphics[width=0.45\textwidth]{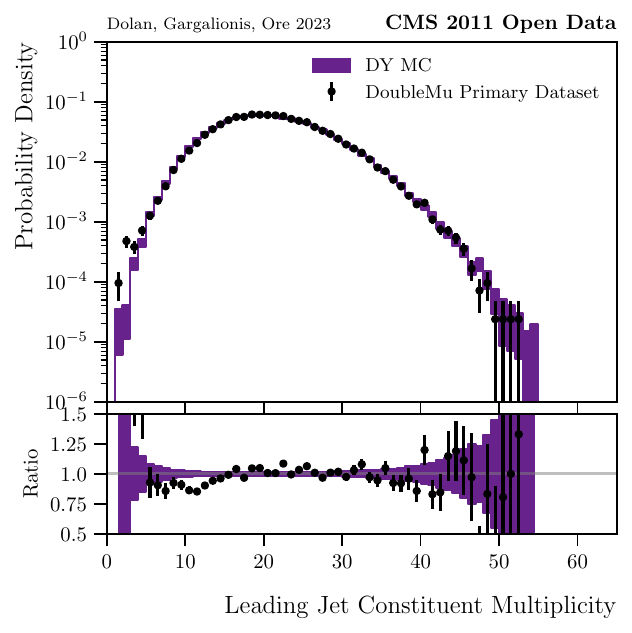}
    }\\[-8pt]
    \subfloat[\label{fig:dijet_sigma_1}]{
        \includegraphics[width=0.45\textwidth]{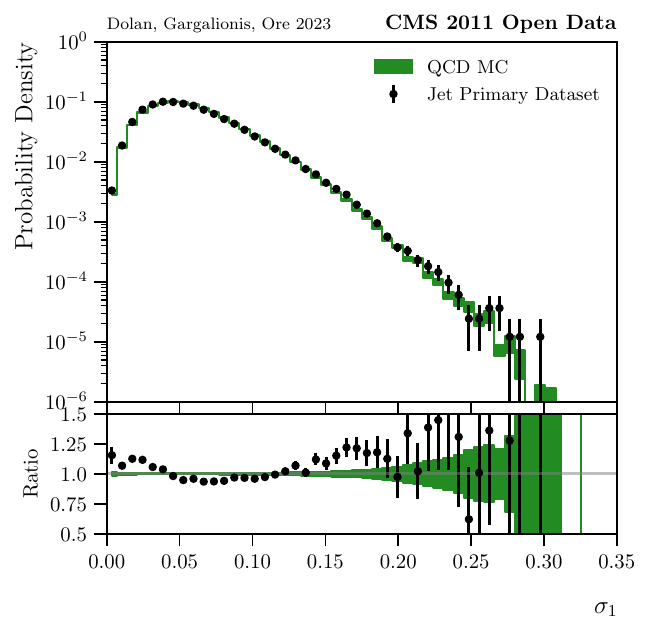}
    }
    \hspace{1cm}
    \subfloat[\label{fig:zjet_sigma_1}]{
        \includegraphics[width=0.45\textwidth]{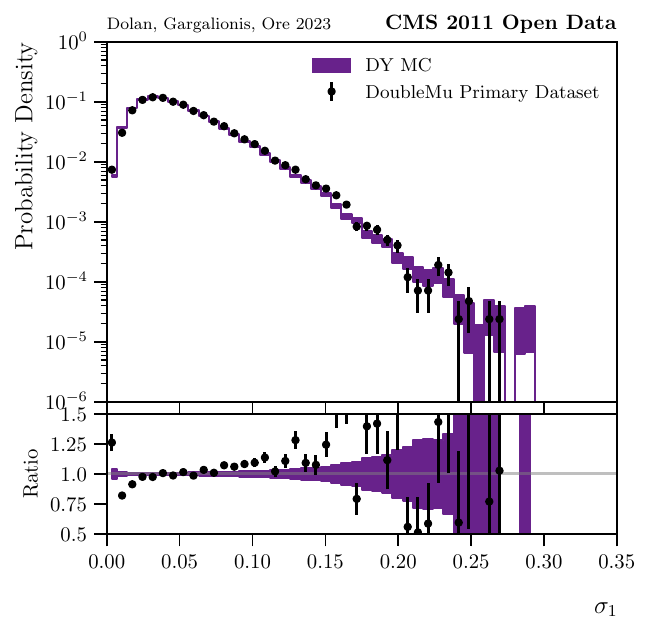}
    }\\[-8pt]
    \subfloat[\label{fig:dijet_ptD}]{
        \includegraphics[width=0.45\textwidth]{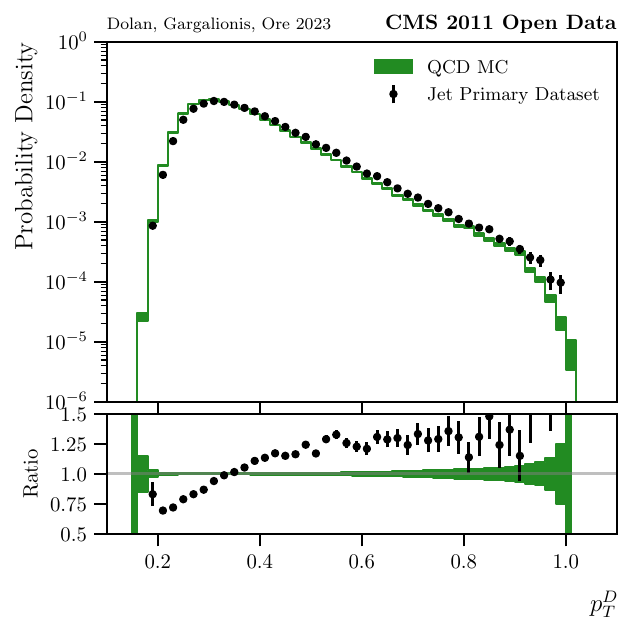}
    }\hspace{1cm}  
    \subfloat[\label{fig:zjet_ptD}]{
        \includegraphics[width=0.45\textwidth]{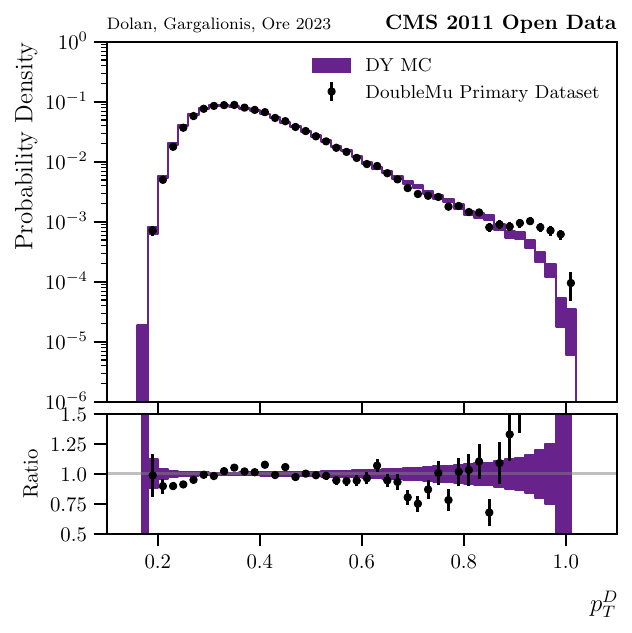}
    }
      
    \caption{Data/MC comparison for jet substructure observables in  dijet events (\protect\subref{fig:dijet_nConstits}, \protect\subref{fig:dijet_sigma_1}, \protect\subref{fig:dijet_ptD}) and \zjet events (\protect\subref{fig:zjet_nConstits}, \protect\subref{fig:zjet_sigma_1}, \protect\subref{fig:zjet_ptD}).}
    \label{fig:additional-substructure}
\end{figure}

Studies of jet physics using the CMS Open Data have also been undertaken in a number of papers by the MIT group. We have compared our processed data with that made publicly available through Ref.~\cite{Komiske:2019jim} in Appendix~\ref{app:mit-comparison}.

\clearpage
\section{Weakly-supervised methods}
\label{sec:weak}

\subsection{Classification without labels}
The classification without labels (CWoLa) paradigm was first explored for high energy physics in~\cite{Metodiev:2017vrx}. In this picture, we assume that two mixtures $M_1$ and $M_2$ are differently proportioned combinations of the same underlying `quark' and `gluon' components such that we can write distributions for any observable $x$ as 
\begin{equation}
    \label{eq:mixture-densities}
    \begin{aligned}
    	p_{M_1}(x) & = f_1 \, p_Q(x) + (1-f_1) \, p_G(x)\, ,\\
    	p_{M_2}(x) & = f_2 \, p_Q(x) + (1-f_2) \, p_G(x)\, ,
    \end{aligned}
\end{equation}
with $f_1$ and $f_2$ being the quark fractions of the mixtures. The Neyman--Pearson lemma states that the optimal binary classifier for the mixtures will be the likelihood ratio $p_{M_1}/p_{M_2}$ which is monotonically related to $p_Q/p_G$. Thus the optimal classifier to discriminate $M_1$ and $M_2$ is also optimal for the underlying quark and gluon components. This means that a neural network trained to classify $M_1$ and $M_2$ (which does not require parton-level labels) should perform as well as a fully-supervised network trained with quark/gluon labels.

In practice, there are various reasons for which a weakly-supervised CWoLa classifier may not be able to match the performance of a fully-supervised classifier. Firstly, the above arguments only guarantee that the two classifiers share an \emph{optimum}. Since neural-network training can only ever approximate the optimal classifier, limitations such as the available training statistics can have impact. Additionally, the framework relies on the assumption that the two mixtures contain the same $p_Q$ and $p_G$, called \emph{sample independence}. Depending on the source of each mixture, there may be some statistical difference between quarks in $M_1$ and those in $M_2$, for example. It is known that soft colour correlations can cause sample dependence for \zjet and dijet samples, although previous studies have observed this to be a small effect, particularly for small-radius jets \cite{Bright-Thonney:2018mxq}. Detector effects may also introduce large differences.\footnote{Detector-induced sample dependence may be alleviated through unfolding methods, as was applied in \cite{Komiske:2022vxg}.} We will assess the level to which sample independence is a valid assumption for our datasets in Section~\ref{sec:results}.

\subsection{Jet topics}
\label{sec:topics}

In order to train a CWoLa classifier, it is not necessary to know the precise quark fractions $f_1$ and $f_2$ of each mixture. It is only required that they are different. However, it is desirable to have a way of measuring these fractions in data as they can provide a comparison point to the predictions of parton-shower programs. Further, knowledge of the fractions allows Eq.\,\ref{eq:mixture-densities} to be inverted, giving an estimate for the component quark/gluon densities:
\begin{equation}
    \label{eq:inverted-distributions}
    \begin{gathered}
        p_{Q}(x) = \frac{(1-f_2)p_{M_1}(x) - (1-f_1)p_{M_2}(x)}{f_1-f_2},\\
        p_{G}(x) = \frac{f_1p_{M_2}(x) - f_2p_{M_1}(x)}{f_1-f_2}.    
    \end{gathered}
\end{equation}
The jet-topics framework~\cite{Metodiev:2018ftz,Komiske:2018vkc,Komiske:2022vxg} provides a method for extracting these fractions. In addition to sample independence of $p_Q$ and $p_G$ this method requires that one can measure an observable in which the distributions are \emph{mutually irreducible}. This further assumption is equivalent to requiring that the support of $p_Q$ is not a subset of the support of $p_G$ and vice-versa. Under these conditions, Eq.\,\ref{eq:inverted-distributions} can be rewritten in terms of \emph{reducibility factors}, defined as
\begin{align}
    \label{eq:red-factors}
    \kappa_{ij}\equiv\kappa(M_i|M_j) = \underset{x}{\operatorname{min}}\,\frac{p_{M_i}(x)}{p_{M_j}(x)}.
\end{align}
Assuming mutual irreducibility of quarks and gluons in $x$, the quark fractions correspond to these $\kappa$ according to
\begin{equation}
    \label{eq:red-fracs}
    f_1 = \frac{1-\kappa_{12}}{1-\kappa_{12}\kappa_{21}},\quad f_2=\kappa_{21}{f_1}\,,
\end{equation}
which allows one to write
\begin{equation}
    \label{eq:demix-distributions}
    \begin{aligned}
        p_{Q}(x) &= \frac{p_{M_1}(x) - \kappa_{12}\,p_{M_2}(x)}{1-\kappa_{12}},\\
        p_{G}(x) &= \frac{p_{M_2}(x) - \kappa_{21}\,p_{M_1}(x)}{1-\kappa_{21}}.
    \end{aligned}    
\end{equation}
Since the $\kappa$ factors are obtainable from samples of $M_1$ and $M_2$, their measurement provides a direct estimate for the quark fractions $f_1,f_2$ and pure distributions that can be performed on data. Importantly, the extracted fractions apply to any observable, not only the $x$ with which the reducibility factors were determined. A number of methods for performing the minimisation of Eq.\,\ref{eq:red-factors} are outlined in~\cite{Komiske:2022vxg}.

The statement that $p_Q(x)$ and $p_G(x)$ are mutually irreducible is equivalent to $\kappa_{QG}=\kappa_{GQ}=0$. If this assumption is not valid for the chosen observable, one can still invert Eq.\,\ref{eq:mixture-densities} to yield fractions and pure distributions so long as the non-zero $\kappa$ factors are known (perhaps from theory or simulation). In this case, Eq.\,\ref{eq:red-fracs} generalises to
\begin{equation}
    \label{eq:eq:red-fracs-general}
    \begin{aligned}
        f_1 &= \frac{1}{1-\kappa_{12}\kappa_{21}}\left(\frac{1-\kappa_{12}}{1-\kappa_{QG}}-\frac{\kappa_{12}\kappa_{GQ}(1-\kappa_{21})}{1-\kappa_{GQ}}\right),\\
        f_2 &= \frac{1}{1-\kappa_{12}\kappa_{21}}\left(\frac{\kappa_{21}(1-\kappa_{12})}{1-\kappa_{QG}}-\frac{\kappa_{GQ}(1-\kappa_{21})}{1-\kappa_{GQ}}\right).
    \end{aligned}
\end{equation}

Of course, categories defined in this manner are no longer completely data-driven. An alternative method for extracting mixture fractions in reducible observables is described in~\cite{Zhu2023MixturePE}. 

\subsection{TopicFlow}

In previous work, some of us developed TopicFlow\,\cite{Dolan:2022ikg}, with which one can train a generative model for the topic distributions $p_Q$ and $p_G$ given estimates of the quark fractions of two mixtures. The loss function of a deep generative model usually takes the form of an expectation value for some function $L$ (e.g.\ the negative log-likelihood in the case of normalising flows) over samples from the target distribution. While one cannot directly sample pure quark or gluon jets from experimental data, we can use Eq.\,\ref{eq:demix-distributions} to rewrite the loss in terms of expectations over mixtures $M_i$. Taking the quark distribution as an example, one has
\begin{equation}
    \label{eq:topicflow-loss}
    \mathcal{L}_Q^\theta = \left\langle L^\theta(x)\right\rangle_{x\sim p_Q} = \left\langle L^\theta(x) \right\rangle_{x\sim p_{M_1}} - \kappa_{12}\,\left\langle L^\theta(x) \right\rangle_{x\sim p_{M_2}},
\end{equation}
where the second equality is true up to normalisation by $1-\kappa_{12}$.

In practice, training models with a loss that balances opposite-sign terms can harm the stability of the optimisation. In the case of the loss in Eq.~\ref{eq:topicflow-loss}, the network may discover that it can `cheat' by ignoring the first term and trying to achieve poor loss on $M_2$, which is often trivial. In this work, we use an alternative loss that \emph{jointly} trains quark and gluon topic models based on Eq.\,\ref{eq:mixture-densities}. Specifically, given a loss function $L$ that depends on the likelihood of the data under the model, we minimise\footnote{For maximum likelihood training, the function $L(x)$ is simply $-\log{x}$. One can of course consider other objective functions including those that depend on the likelihood only implicitly.}
\begin{equation}
    \label{eq:topicflow2-loss}
	\mathcal{L}_\mathrm{TF} = \big\langle L[q_1^{\theta_Q,\theta_G}(x_1)]\big\rangle_{x_1\sim M_1} + \big\langle L[q_2^{\theta_Q,\theta_G}(x_2)]\big\rangle_{x_2\sim M_2},
\end{equation}
where each $q_i^{\theta_Q,\theta_G}$ is a likelihood parameterised by two networks according to 
\begin{equation}
    \begin{aligned}
    	q_1^{\theta_Q,\theta_G}(x) &= f_1q^{\theta_Q}(x) + (1-f_1)q^{\theta_G}(x),\\
    	q_2^{\theta_Q,\theta_G}(x) &= f_2q^{\theta_Q}(x)+ (1-f_2)q^{\theta_G}(x).
    \end{aligned}    
\end{equation}

In contrast to Eq.\,\ref{eq:topicflow-loss}, the loss in Eq.\,\ref{eq:topicflow2-loss} is convex and therefore facilitates stable training. We find that models trained with this objective yield better likelihoods and generate higher quality samples.

While it is straightforward to apply Eq.\,\ref{eq:demix-distributions} to histograms of two mixed datasets, generative models offer a number of benefits compared to binned approaches, particularly when $x$ is a high-dimensional observable. Firstly, generative models allow oversampling to smooth statistics over small datasets. In some cases, subject to the available data and the inductive bias of the model, deep generative models have been observed to contain more information than the data on which they were trained~\cite{Butter:2020qhk, Bieringer:2022cbs}. A second advantage is the possibility of conditioning the generative model on auxiliary inputs such that distributions can be interpolated between measured points (e.g.~\cite{Hallin:2021wme,Algren:2023qnb}). The use case we will explore in this work, however, is \emph{generative classification} where one constructs a quark/gluon likelihood ratio using individually modelled distributions.

In \cite{Dolan:2022ikg}, TopicFlow was applied only to simulated data. However it is compatible with real data since the underlying model (a normalising flow) is unsupervised. This study therefore represents the first application of the method to real jets.

\section{Training procedure and network architecture}
\label{sec:training}

To conduct our comparison between the CWoLa and fully-supervised paradigms we consider a number of classifiers, which we define according their training datasets as outlined in Table~\ref{tab:dataset-definition}. To enable a comparison independent of dataset size, we partition the Data CWoLa dataset into 90\%/10\% training/testing splits and then match the training splits for the MC datasets to this size. As a result, we have more statistics in the MC testing splits. To account for the class imbalance (all datasets contain more gluon-like jets than quark-like\footnote{For the mixed datasets, this is reflected directly in Table~\ref{tab:dataset-sizes}. For the pure dataset the imbalance comes from the fact that the QCD MC is gluon-enriched.}), we train models with class weights. In practice, we found no difference in performance compared to truncating the majority class to match the minority class.

\begin{table}
    \centering
    \begin{tabular}{rr}
        \toprule
        {\bf Model name} & \textbf{Classes}    \\
        \midrule
        Fully supervised & Quarks and gluons from QCD MC \\ MC CWoLa & DY MC and
        QCD MC                                           \\ Data CWoLa & {\tt DoubleMu} and {\tt Jet} \\
        \bottomrule
    \end{tabular}
    \caption{Outline of the datasets on which each of the models was trained.}
    \label{tab:dataset-definition}
\end{table}

As a representation for the jets, we make use of Energy Flow Polynomials (EFPs)~\cite{Komiske:2017aww}, defined in Appendix~\ref{app:substructure-definitions}. EFPs form an over-complete basis for infrared and collinear safe (IRC) jet observables and are thus suited for use with real data, since physical detectors have finite energy/angle resolution. It was also observed in~\cite{Bright-Thonney:2018mxq} that IRC-safe observables exhibit smaller sample dependence.
We consider connected EFPs with $\beta=\frac{1}{2}$ and number of edges~$d\in[1,3\,]$.
\begin{figure}
    \centering
    \subfloat[\label{efp_1_dijet}]{
        \includegraphics[width=0.45\textwidth]{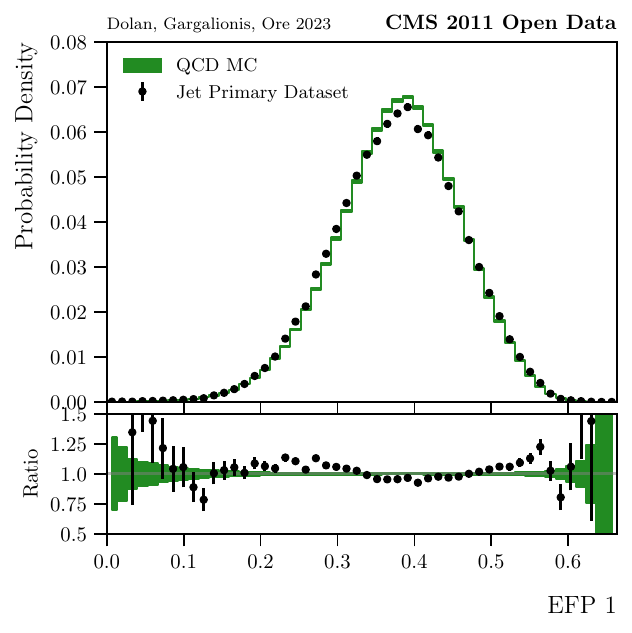}
    }\hfill
    \subfloat[\label{efp_2_dijet}]{
        \includegraphics[width=0.45\textwidth]{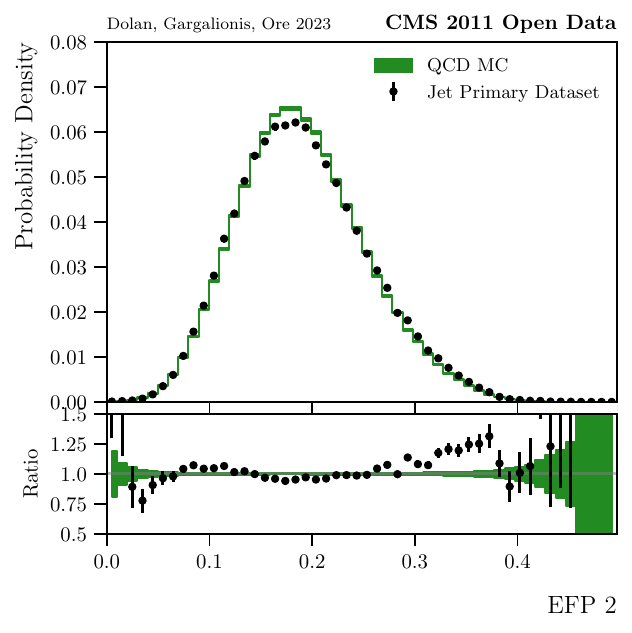}
    }\\
    \subfloat[\label{efp_1_zjet}]{
        \includegraphics[width=0.45\textwidth]{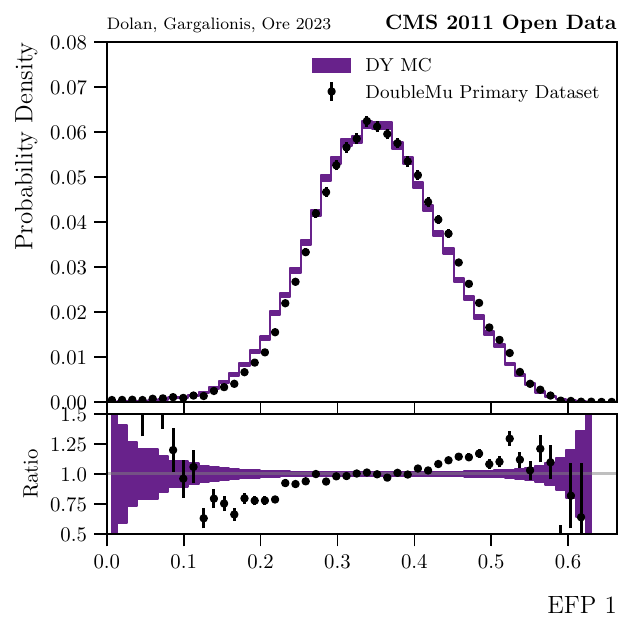}
    }\hfill
    \subfloat[\label{efp_2_zjet}]{
        \includegraphics[width=0.45\textwidth]{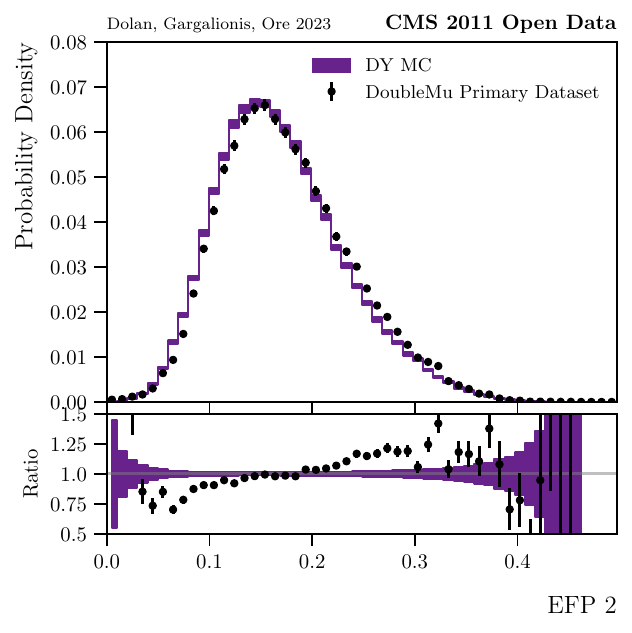}
    }
    \caption{Data/MC comparison for two Energy Flow Polynomials in \protect\subref{efp_1_dijet}, \protect\subref{efp_2_dijet} dijet events and \protect\subref{efp_1_zjet}, \protect\subref{efp_2_zjet} \zjet events.}
    \label{fig:efps}
\end{figure}
Figure~\ref{fig:efps} shows histograms for two of the eight total EFPs in the basis, corresponding to the graphs
\begin{equation}
    \text{EFP 1} \equiv \vcenter{\hbox{\includegraphics[width=2cm]{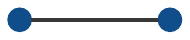}}}, \qquad
    \text{EFP 2} \equiv \vcenter{\hbox{\includegraphics[width=2cm]{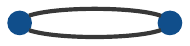}}}.
\end{equation}
The distributions exhibit small data-MC discrepancies, in particular systematically smaller EFP values in DY simulation compared to the \texttt{DoubleMu} primary dataset. As with the other substructure observables shown in Figure~\ref{fig:additional-substructure} above, the error bands are statistics-only.

To stabilise the network training, we apply some preprocessing steps to the EFPs. These include a log-scaling followed by a translation such that the mean value is zero. We then rotate to the principal components basis before scaling the data such that the standard deviation is one. The transformations are determined from the training split of the {\tt DoubleMu} and {\tt Jet} datasets and reused for all other datasets.

\begin{figure}
    \centering
    \subfloat[\label{fig:efp_1_sample_dependence}]{
        \includegraphics[width=0.33\textwidth]{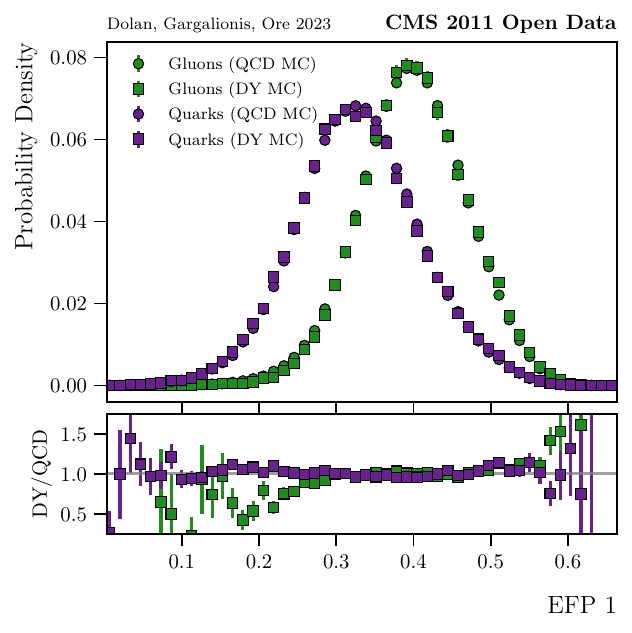}
    }
    \subfloat[\label{fig:efp_2_sample_dependence}]{
        \includegraphics[width=0.33\textwidth]{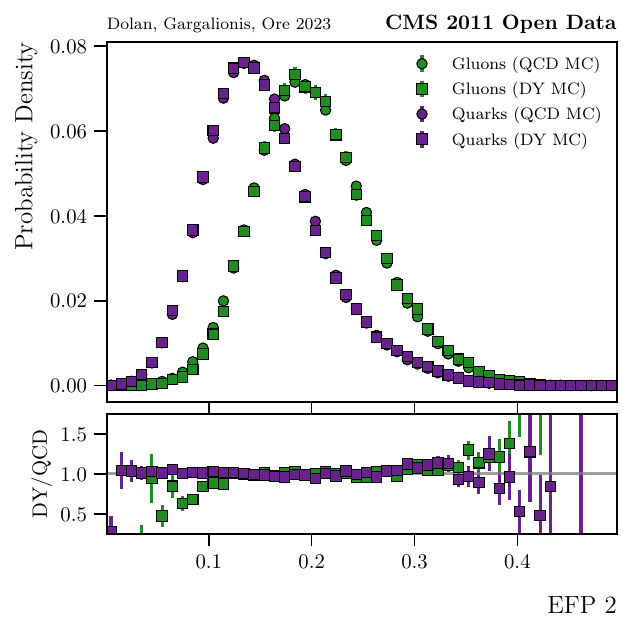}
    }\\
    \subfloat[\label{fig:mixed_preds}]{
    \includegraphics[width=0.33\textwidth]{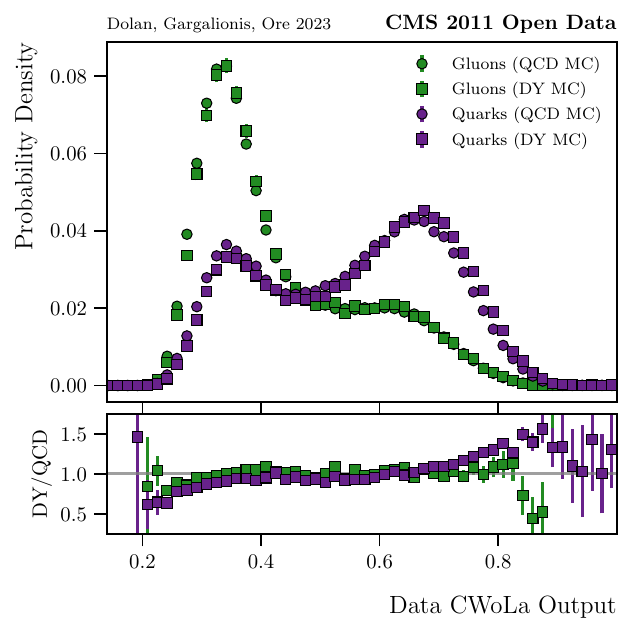}
    }
    \subfloat[\label{fig:mcmixed_preds}]{
    \includegraphics[width=0.33\textwidth]{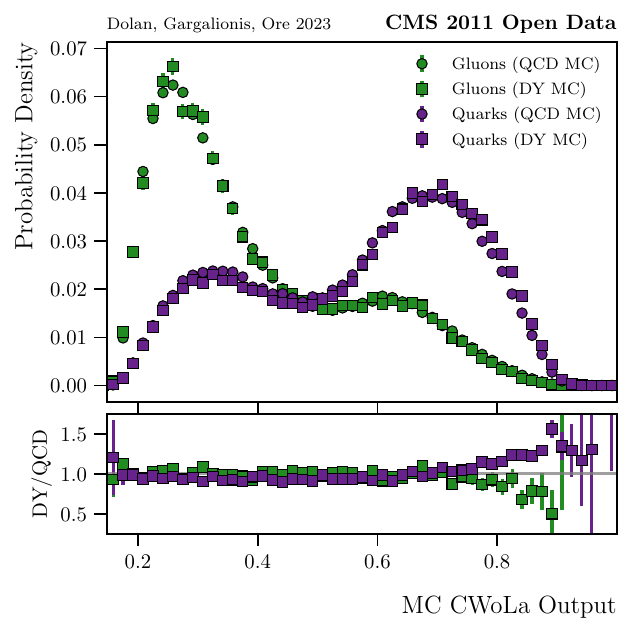}
    }
    \subfloat[\label{fig:pure_preds}]{
    \includegraphics[width=0.33\textwidth]{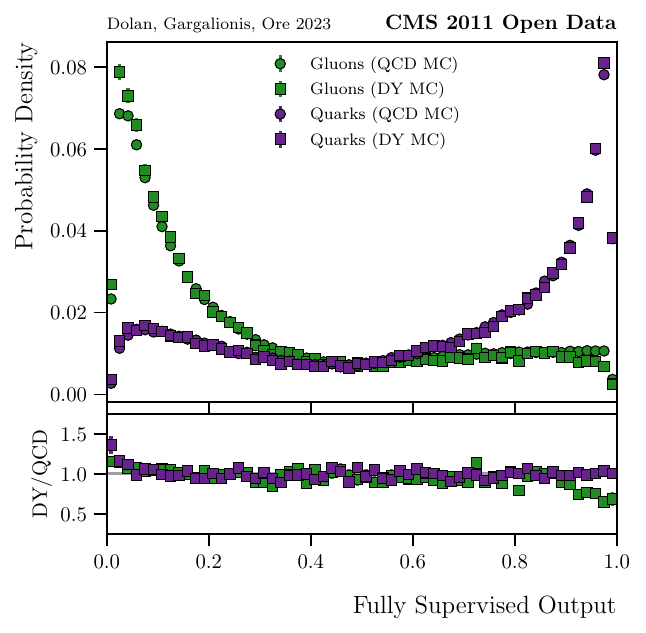}
    }
    \caption{Histograms of the first two Energy Flow Polynomials (\protect\subref{fig:efp_1_sample_dependence}, \protect\subref{fig:efp_2_sample_dependence}) and classifier predictions (\protect\subref{fig:mixed_preds}, \protect\subref{fig:mcmixed_preds}, \protect\subref{fig:pure_preds}) for quarks and gluons in dijet and \zjet events from CMS simulation. The agreement among similar colours illustrates approximate sample independence in the datasets.}
    \label{fig:sample-dependence}
\end{figure}

For the classification networks, we use fully-connected multilayer perceptrons (MLPs) consisting of five layers with 128 nodes each and RELU activations. We apply dropout at each layer at a rate 0.1 to control overfitting. The networks are trained to minimise the binary cross-entropy using the Adam optimiser with initial learning rate $10^{-4}$ and batch size 500. When training on simulated datasets, we include the MC weights when aggregating losses over a batch. The loss on a validation set (10\% the size of the training set) is monitored at each epoch and the learning rate is reduced by a factor 10 if no improvement is seen for 5 epochs. If the validation loss has not improved for 10 epochs or 100 total epochs have been completed, the training is halted and the model weights from the best epoch are restored.

For TopicFlow, we train continuous normalising flows with the Flow Matching objective of \cite{Lipman2022FlowMF}. To adapt this loss according to Eq.\,\ref{eq:topicflow2-loss}, we replace the single regression network with a linear combination of `quark' and `gluon' networks as $v_\theta\to f_iv_{\theta_Q} + (1-f_i)v_{\theta_G}$.\footnote{Interestingly, for two vector fields $v_Q$ and $v_G$ respectively generating probability paths $p_{t,Q}$ and $p_{t,G}$ it does \emph{not} follow that the combination $fv_Q+(1-f)v_G$ generates $fp_{t,Q}+(1-f)p_{t,G}$. Despite this, we find empirically that models trained in this way learn the expected topic distributions.} For each regression network we use an MLP consisting of three 512-node layers with GELU activations and a skip connection between the first and third layer. Dropout is applied after each layer at a rate of 0.2. We use the Adam optimiser with an initial learning rate of $10^{-3}$ which is decayed by a factor 10 if there is no improvement for 20 epochs. Training is halted after 200 epochs, or once the learning rate has decayed below $10^{-6}$.

\section{Results}
\label{sec:results}

\subsection{Extracted quark fractions}
\label{sec:fractions}

Here we investigate the quark/gluon content of each of the datasets using the jet topics. It is first necessary, however to assess whether or not the key assumptions of the framework are valid for our datasets. To validate the assumption of sample independence, we plot in Figure~\ref{fig:sample-dependence} the MC-matched quark and gluon distributions of two EFPs and the outputs of one of each type of classifier, both for DY and QCD simulation. In each observable, there are small but noticeable differences between the different MC datasets. However, the general shape of the distributions agree to a good approximation, with deviations mostly residing in the tails. In the EFP distributions, shown in the top row, the quark distributions are more similar across the \zjet and dijet samples compared to the gluon distributions. Interestingly, when considering the outputs of the two CWoLa models in Figs.\,\ref{fig:mixed_preds} and \ref{fig:mcmixed_preds} (which depend only on the 8 EFPs) the opposite trend is evident. Specifically, both of these models assign slightly higher scores to DY MC quarks than QCD MC quarks, as indicated by the uptick of purple squares in the DY/QCD ratio. The predictions of the CWoLa models exhibit the greatest sample dependence of the observables we consider, with deviations up to 25\% in the worst regions. Meanwhile, the fully-supervised network has outputs that are relatively consistent across the MC samples.

A fully-fledged measurement of any quantity requires a robust estimate of systematic uncertainties. In this essentially exploratory paper we have not attempted an analysis of these. All uncertainties presented are statistical. Consequently we refer to any quantities we derive in this paper as estimates and not as measurements.

To extract reducibility factors and quark fractions in each dataset, we train classifiers and apply the receiver operating characteristic (ROC) curve fit method (described in~\cite{Komiske:2022vxg}) to their outputs. This involves fitting the curve $\big(\varepsilon_{M_1}(t), \varepsilon_{M_2}(t)\big)$, where
\begin{equation}
    \varepsilon_M(t) = \frac{1}{\left|M\right|}\sum_{x\in M}\Theta(c(x)-t)
\end{equation}
is the efficiency of classifier~$c$ in dataset~$M$ at threshold~$t$. The gradients at the (0,0) and (1,1) end points of this curve correspond to $\kappa_{21}$ and $1/\kappa_{12}$ respectively. For each type of classifier (per Table \ref{tab:dataset-definition}), we perform such extractions with 10 independently trained networks on both the primary and simulated datasets. Note that for these estimates we evaluate the models on the full datasets, including the training split.

\begin{figure}
    \centering
    \subfloat[\label{fig:pure_reducibilities}]{
        \includegraphics[width=0.33\textwidth]{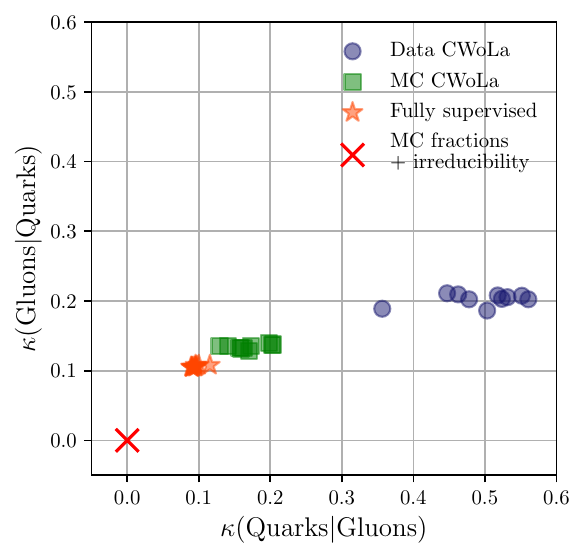}
    }
    \subfloat[\label{fig:mcmixed_reducibilities}]{
        \includegraphics[width=0.33\textwidth]{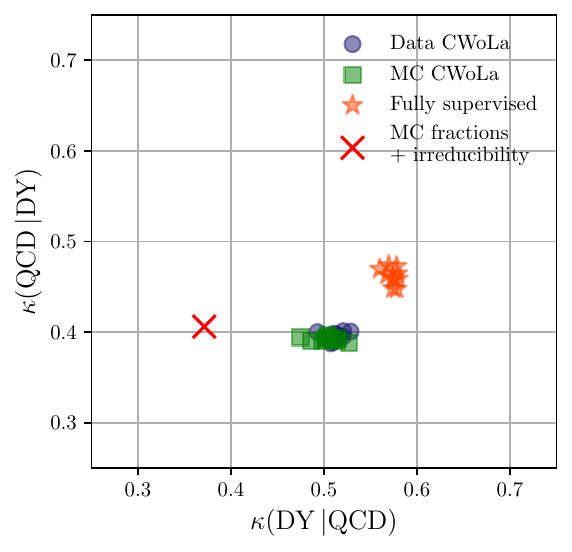}
    }
    \subfloat[\label{fig:mixed_reducibilities}]{
        \includegraphics[width=0.33\textwidth]{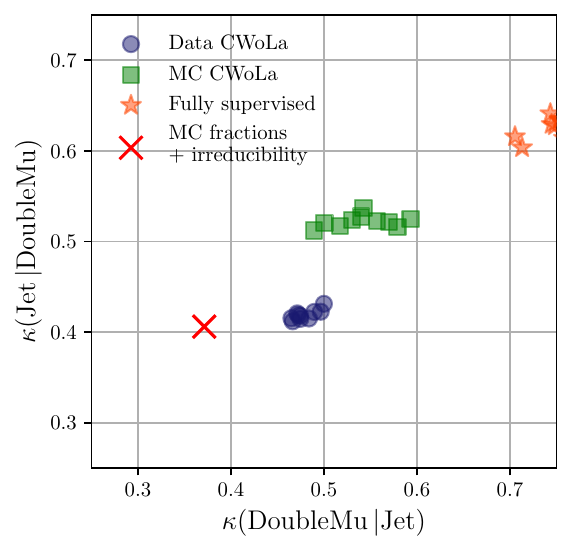}
    }    
    \caption{Extracted reducibility factors for classifier outputs in each of the datasets \protect\subref{fig:pure_reducibilities} Quark/Gluon simulation, \protect\subref{fig:mcmixed_reducibilities} DY/QCD simulation, \protect\subref{fig:mixed_reducibilities} \texttt{DoubleMu}/\texttt{Jet} primary. The red cross indicates the expectation for a mutually irreducible observable given fractions determined by MC parton labels.}
    \label{fig:reducibilities}
\end{figure}

Fig.\,\ref{fig:reducibilities} shows the extracted reducibilities between the quark- and gluon-enriched categories in each dataset. We also indicate with a red cross the expected result for a mutually irreducible observable given that the quark fractions of our MC samples are
\begin{equation}
    \label{eq:mc-fractions}
    f_\text{DY} = 0.740,\ f_\text{QCD} = 0.301,
\end{equation}
as determined by summing parton labels including weights, and $(f_\text{Quarks},f_\text{Gluons}) = (1,0)$ by definition. Since we have observed that the quark and gluon distributions are sample independent to a good approximation in MC, we can attribute disagreement between the extractions and the MC predictions to some level of mutual \emph{reducibility}. In particular, the non-zero $\kappa$'s in Fig.\,\ref{fig:pure_reducibilities} indicate that quarks and gluons in QCD MC are not irreducible in the output of the classifiers. The fully-supervised networks have reducibilities closest to zero, which is consistent with being the strongest quark/gluon discriminator in MC (classification metrics for each model will be presented in Sec.\,\ref{sec:classification}). In principle, one could use these estimates of $\kappa_{QG}$ and $\kappa_{GQ}$ to estimate quark fractions with these networks via Eq.\,\ref{eq:eq:red-fracs-general}.

In Fig.\,\ref{fig:mcmixed_reducibilities} we again see a mismatch between the extracted $\kappa$'s and those expected under mutual irreducibility, shown by the red cross. Interestingly, however, the extractions from the CWoLa models do agree closely for \redkappa{QCD}{DY}, indicating that the one-way irreducibility condition $\kappa_{GQ}=0$ is a valid assumption.\footnote{In fact the extractions slightly underestimate \redkappa{QCD}{DY} which cannot be explained by reducibility --- a non-zero $\kappa_{GQ}$ causes an overestimate of $\kappa_{21}$.} Naively this seems to disagree with the results in Fig.~\ref{fig:pure_reducibilities}, which indicate a non-zero $\kappa_{GQ}$ for both CWoLa networks. This could be due to sample dependence since the left panel does not include the DY simulated dataset. The other possibility is that this a statistical effect due to the fact that the reducibility factors depend on the endpoints of the ROC curves, which are sparsely populated for pure datasets.

\begin{table}[]
    \centering
    \begin{tabular}{lcc}
    \toprule
    {\bf Classifier} & $\bm{\kappa_{QG}}$ & $\bm{\kappa_{GQ}}$\\
    \midrule
    Data CWoLa & $0.169\pm0.012$ & - \\
    MC CWoLa   & $0.157\pm0.018$ & - \\
    Fully Supervised & $0.249\pm0.008$ & $0.0711\pm0.011$\\
    \bottomrule
    \end{tabular}
    \caption{Quark/gluon reducibility factors of each classifier required to explain the measured \redkappa{DY}{QCD} and \redkappa{QCD}{DY} in Fig.\,\ref{fig:mcmixed_reducibilities} given the known MC quark fractions. Uncertainties are calculated as the standard deviation over ten networks.}
    \label{tab:qg_reducibilities}
\end{table}

We can obtain the values of $\kappa_{QG}$ and $\kappa_{GQ}$ that rectify disagreements in \redkappa{DY}{QCD} or \redkappa{QCD}{DY} between the measured values and expected values assuming irreducibility, with
\begin{equation}
    \begin{gathered}
        \kappa_{QG} = 1-\frac{1-\redkappa{DY}{QCD}}{f_\text{DY}-\redkappa{DY}{QCD}f_\text{QCD}}\,,\\
        \kappa_{GQ} = \frac{\redkappa{QCD}{DY}f_\text{DY} - f_\text{QCD}}{1 - f_\text{QCD} - \redkappa{QCD}{DY}(1-f_\text{DY})}\,.
    \end{gathered}    
\end{equation}
We calculate these factors using the ten model points for each type of classifier in Fig.~\ref{fig:mcmixed_reducibilities} using the MC fractions in Eq.\,\ref{eq:mc-fractions}. The results are shown in Tab.\,\ref{tab:qg_reducibilities} which reports the mean $\pm$ standard deviation across ten classifiers of the given type. We exclude the $\kappa_{GQ}$ values for both CWoLa networks since they are already consistent with $\kappa_{GQ} = 0$. While the value of $\kappa_{QG}$ for MC CWoLa agrees well with the average value of $\kappa_{QG}$ for the green boxes in Fig.~\ref{fig:pure_reducibilities}, the remaining estimates do not. This could be another avatar of sample dependence.

Finally, Fig.\,\ref{fig:mixed_reducibilities} shows the measured reducibilities of the primary datasets under each type of classifier. While we also show the MC expectation in this panel, in this case it is only for reference since we do not know the true quark fractions in the {\tt DoubleMu} or {\tt Jet} datasets. However, even without knowing  $f_\mathrm{DoubleMu}$ or $ f_\mathrm{Jet}$, the reducibilities still yield information on the classification strength. Specifically, the fact that data CWoLa predictions have the smallest reducibilities in the {\tt DoubleMu} and {\tt Jet} categories indicates it is the most powerful of the three classifiers.

\begin{figure}
    \centering
    \subfloat[\label{fig:mcmixed_fractions}]{
        \includegraphics[width=0.33\textwidth]{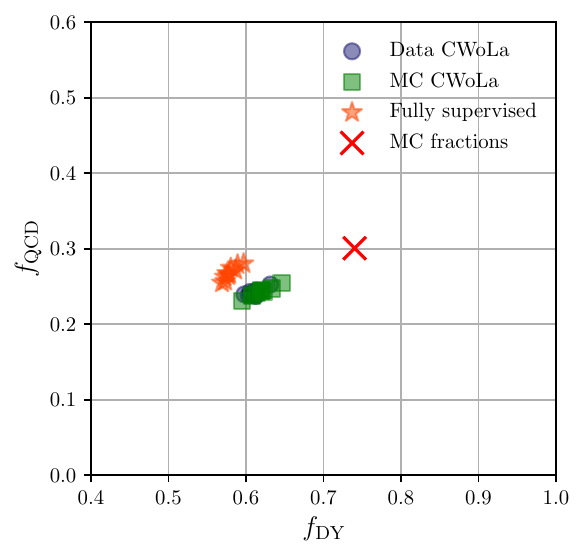}
    }\hspace{1cm}
    \subfloat[\label{fig:mixed_fractions}]{
        \includegraphics[width=0.33\textwidth]{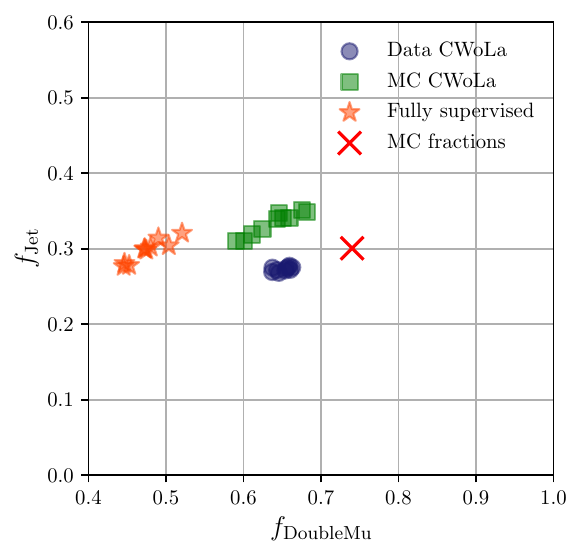}
    }

    \subfloat[\label{fig:mcmixed_fractions_corrected}]{
        \includegraphics[width=0.33\textwidth]{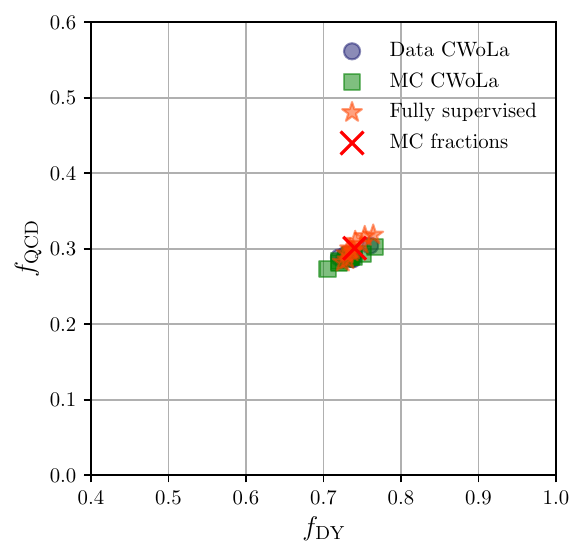}
    }\hspace{1cm}
    \subfloat[\label{fig:mixed_fractions_corrected}]{
        \includegraphics[width=0.32\textwidth]{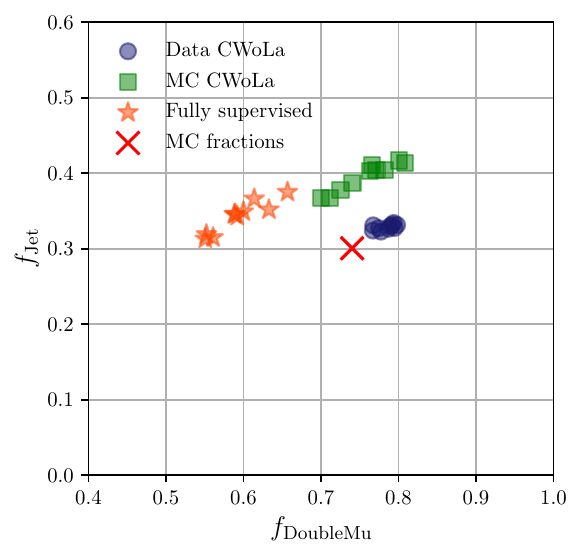}
    }    
    \caption{Quark fractions for DY / QCD simulated datasets (\protect\subref{fig:mcmixed_fractions}, \protect\subref{fig:mcmixed_fractions_corrected}) and \texttt{DoubleMu} and \texttt{Jet} primary datasets (\protect\subref{fig:mixed_fractions}, \protect\subref{fig:mixed_fractions_corrected}) according to jet topics extractions using each type of classifier compared to the parton level predictions. The top row assumes $\kappa_{QG}=\kappa_{GQ}=0$ (mutual irreducibility) for the network outputs  whereas the bottom row assumes non-zero $\kappa$'s such that the correct fractions are recovered in MC.}
    \label{fig:fractions}
\end{figure}

We now estimate the quark fractions of the mixed datasets. As discussed in Sec.\,\ref{sec:topics}, if we assume the quark and gluon categories are mutually irreducible, then the fractions depend only on the reducibilities between $M_1$ and $M_2$ as in Eq.\,\ref{eq:red-fracs}. Otherwise, the quark/gluon reducibility factors must be included per Eq.\,\ref{eq:eq:red-fracs-general}. The top row of Fig.\,\ref{fig:fractions} shows the fractions evaluated in the former fashion, assuming $\kappa_{QG}=\kappa_{GQ}=0$. In both the MC (\ref{fig:mcmixed_fractions}) and primary (\ref{fig:mixed_fractions}) datasets, the extractions are different from the MC prediction shown by the red cross. When considering the primary datasets, the MC prediction is only a reference and the disagreement could be explained by mismodelling of quarks and gluons in simulation. Taking the Data CWoLa extraction as the representative estimate, we obtain quark fractions in the primary datasets of
\begin{equation}
    \label{eq:irreducible_extraction}
    f_\mathrm{DoubleMu}=0.651\pm0.009,\ f_\mathrm{Jet} = 0.273\pm0.003,
\end{equation}
where uncertainties are calculated as the standard deviation over the ten points.

\begin{table}[]
    \centering
    \begin{tabular}{ccc}
    \toprule
    {\bf Label} & $\bm{f_\mathrm{DoubleMu}}$ & $\bm{f_\mathrm{Jet}}$\\
    \midrule
    $S$ & 0.740 & 0.301 \\
    $T$ & $0.651\pm0.009$ & $0.273\pm0.003$ \\
    $R$ & $0.784\pm0.011$ & $0.329\pm0.003$ \\
    \bottomrule
    \end{tabular}
    \caption{Summary of our three estimates for the quark fractions of the {\tt DoubleMu} and {\tt Jet} datasets. $S$ (simulation) is taken directly from the sum of MC weights (Eq.\,\ref{eq:mc-fractions}). $T$ (topic) and $R$ (reducible) are extracted using Data CWoLa classifiers as described in Sec.\,\ref{sec:fractions}. $T$ assumes quarks and gluons are mutually irreducible in the classifier score whereas $R$ assumes the same $\kappa_{QG}$ as MC (given in Table\,\ref{tab:qg_reducibilities}). The reported uncertainties are calculated as the standard deviation over the ten points.}
    \label{tab:fraction-summary}
\end{table}

For the MC datasets, however, the discrepancy between the red cross and the CWoLa extractions can only be explained by the quark and gluon categories having reducible distributions in the network outputs. In the bottom row of Fig.\,\ref{fig:fractions} we repeat the estimates using the MC-derived quark/gluon reducibilities from Tab.\,\ref{tab:qg_reducibilities}, effectively calibrating the extractions using known fractions in MC. For the simulated datasets in Fig.\,\ref{fig:mcmixed_fractions_corrected}, the measured fractions now all align closely with the MC prediction as expected. In Fig.\,\ref{fig:mixed_fractions_corrected}, we see that the same corrections applied to the {\tt DoubleMu} and {\tt Jet} datasets shift all extracted quark fractions toward larger values.\footnote{This measurement is akin to a template fit to the data given quarks and gluons in MC.} The Data CWoLa estimates now lie beyond the MC prediction compared to when mutual irreducibility was assumed. Taking the average of these estimates for Data CWoLa gives:
\begin{equation}
    \label{eq:reducible_extraction}
    f_\mathrm{DoubleMu}=0.784\pm0.011,\ f_\mathrm{Jet} = 0.329\pm0.003.
\end{equation}
Again, uncertainties are the standard of ten points. While the above result assumes that the reducibility of $p_Q$ and $p_G$ in data is the same as in simulation, the result in Eq.\,\ref{eq:irreducible_extraction} equally relies on assuming a value for $\kappa_{QG}$ (0 in that case). It is interesting that the `$\kappa$-corrected' predictions in Fig.\,\ref{fig:mixed_fractions_corrected} do not reproduce the expected $f_\mathrm{DoubleMu}, f_\mathrm{Jet}$ based on MC parton labels. This leads one to conclude that either $(f_\mathrm{DoubleMu}, f_\mathrm{Jet})$ or $\kappa_{QG}$ is not well modelled in MC, assuming sample dependence does not play a role. In the absence of another means to estimate the quark/gluon reducibilities in data, we will present results using each of the possible fractions. For clarity, we assign labels of $S$ (simulation), $T$ (topic) and $R$ (reducible) respectively to Eqs.\,\ref{eq:mc-fractions}, \ref{eq:irreducible_extraction}, and \ref{eq:reducible_extraction} as summarised in Table\,\ref{tab:fraction-summary}. Note that the question of reducibility is only relevant for extracting quark fractions, and the following analysis only requires that sample independence holds.

\subsection{Topic distributions}

Here we present the results of TopicFlow trained on the \texttt{DoubleMu} and \texttt{Jet} datasets.
\begin{figure}
    \subfloat[\label{fig:d3_quark_efp_0_parton_topicflow}]{
        \includegraphics[width=0.45\textwidth]{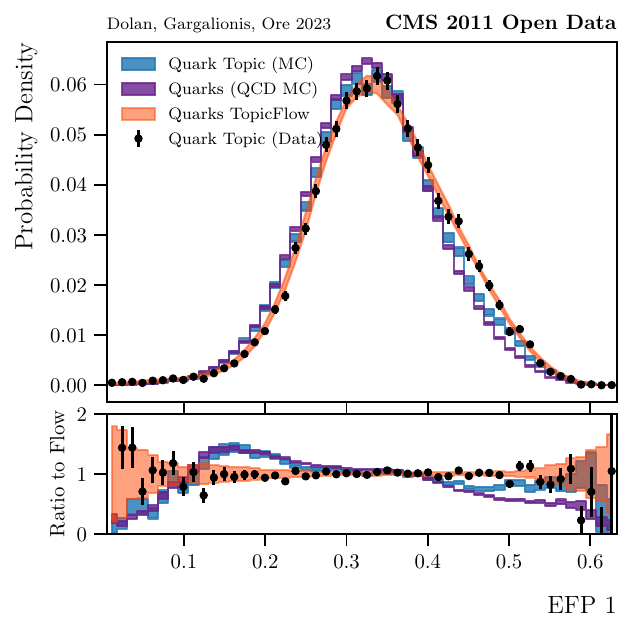}
    }\hfill
    \subfloat[\label{fig:d3_quark_efp_1_parton_topicflow}]{
        \includegraphics[width=0.45\textwidth]{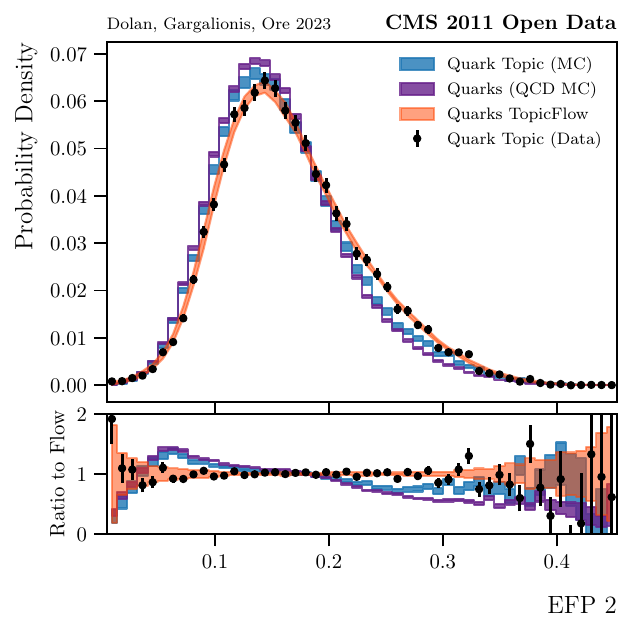}
    }\\
    \subfloat[\label{fig:d3_gluon_efp_0_parton_topicflow}]{
        \includegraphics[width=0.45\textwidth]{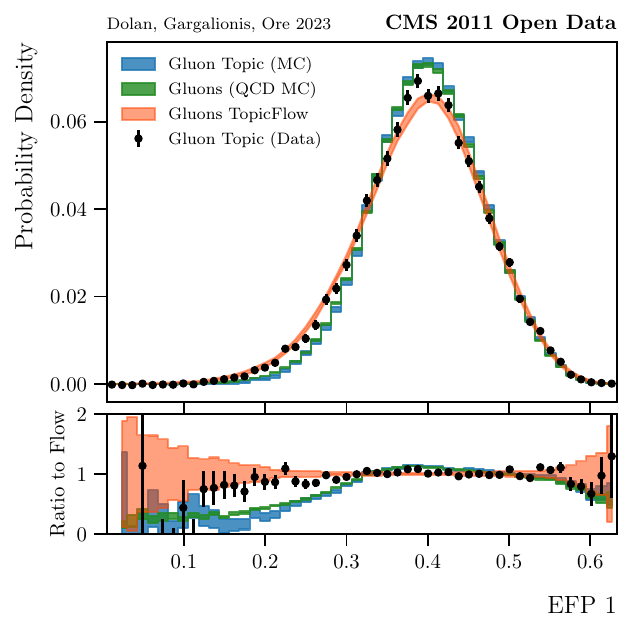}
    }\hfill
    \subfloat[\label{fig:d3_gluon_efp_1_parton_topicflow}]{
        \includegraphics[width=0.45\textwidth]{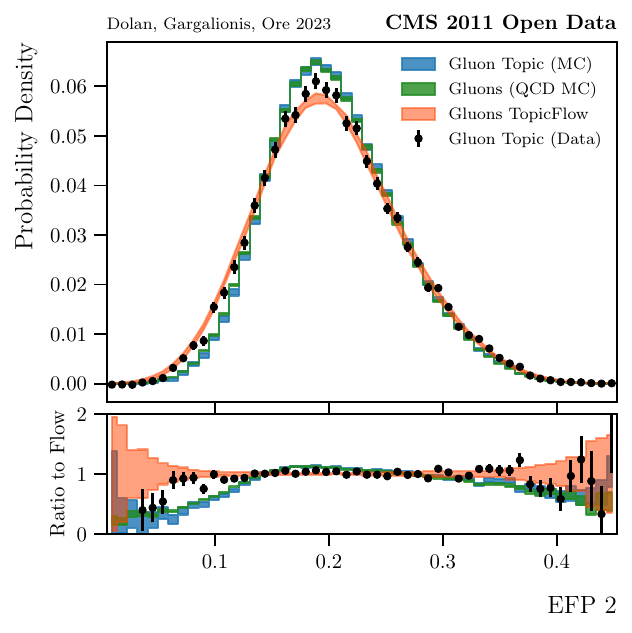}
    }
    \caption{Distribution of Energy Flow Polynomials for quarks (\protect\subref{fig:d3_quark_efp_0_parton_topicflow}, \protect\subref{fig:d3_quark_efp_1_parton_topicflow}) and gluons (\protect\subref{fig:d3_gluon_efp_0_parton_topicflow}, \protect\subref{fig:d3_gluon_efp_1_parton_topicflow}) according to QCD MC and jet topics constructed from histograms of the primary datasets and TopicFlow. All topic distributions assume fractions determined from MC parton labels (Eq.\,\ref{eq:mc-fractions}). Bands show the mean $\pm$ one standard deviation over 10 flow models.}
    \label{fig:topicflow}
\end{figure}
Fig.\,\ref{fig:topicflow} shows the results for quark and gluon topics, comparing both to `pure' predictions from QCD simulation and the jet topic distributions formed by subtracting histograms of the mixed datasets according to Eq.\,\ref{eq:demix-distributions}. For best comparison with MC, all topic distributions (and TopicFlow loss) use fractions~$S$. We also trained models assuming fractions $T$ and $R$, although the plots are qualitatively the same in each case. From each TopicFlow model, we generate 200k samples and the band shows the mean $\pm$ one standard deviation across ten models. Note that the samples are generated in the preprocessed basis, so the EFPs are recovered by inverting these transformations.

The first observation is that while the topic distributions constructed in MC mostly agree with the pure QCD predictions, neither MC distribution closely matches the topics constructed in data. The data-MC discrepancies are in fact larger than were displayed in Fig.~\ref{fig:efps} for the mixed simulation. This is despite using fractions~$S$, which assume the same mixture proportions as MC. As discussed previously, it is possible that the data-MC difference is covered by systematics that we cannot quantify in this study.
On the other hand, the TopicFlow networks closely follow the quark/gluon topics in data, showing no large biases in the bulk of the distributions. We verified that the same results are found when comparing to DY simulation.

\subsection{Classification}
\label{sec:classification}

\begin{figure}
    \centering
    \subfloat[\label{fig:d3_mc_roc}]{
        \includegraphics[width=0.45\textwidth]{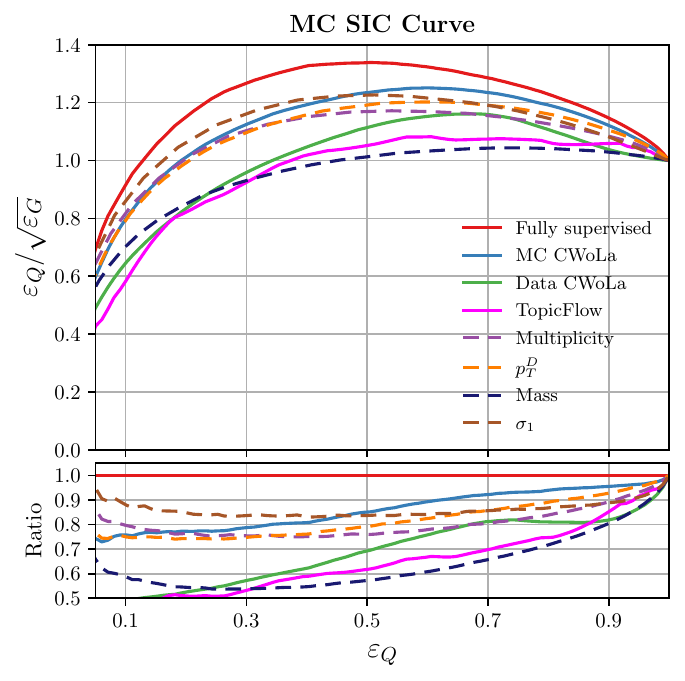}
    }\hfill
    \subfloat[\label{fig:d3_parton_roc}]{
        \includegraphics[width=0.45\textwidth]{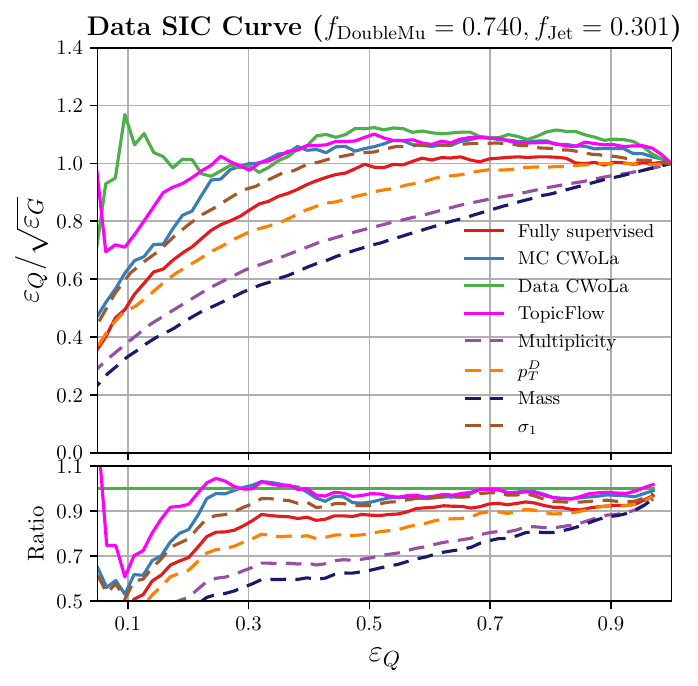}
    }\\
    \subfloat[\label{fig:d3_cwola_roc}]{
        \includegraphics[width=0.45\textwidth]{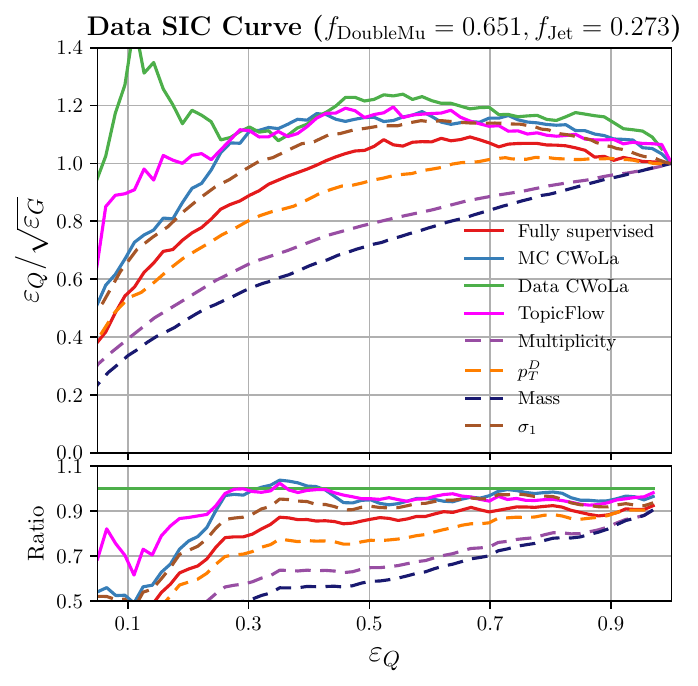}
    }\hfill
    \subfloat[\label{fig:d3_cwola_corrected_roc}]{
        \includegraphics[width=0.45\textwidth]{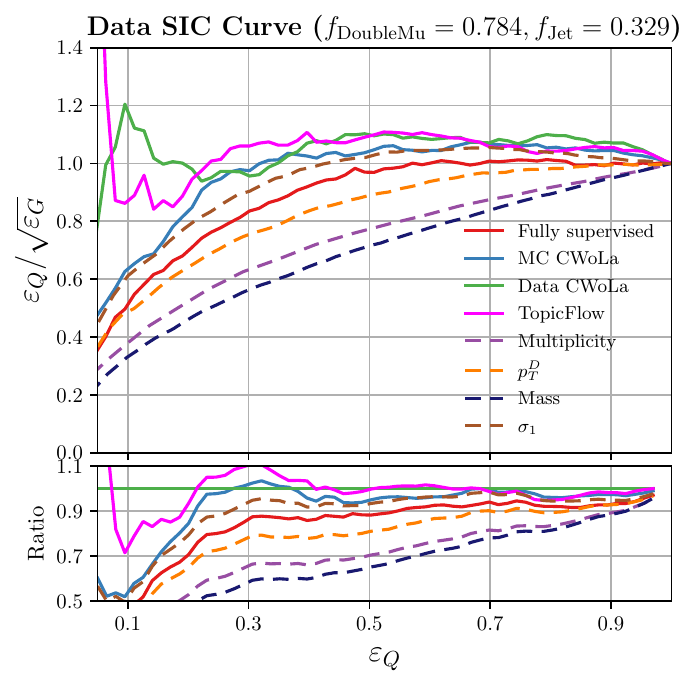}
    }
    \caption{SIC curves constructed using \protect\subref{fig:d3_mc_roc} QCDSim parton labels and the Primary datasets via Eq.\,\ref{eq:data-roc} with \protect\subref{fig:d3_parton_roc} fractions $S$, \protect\subref{fig:d3_cwola_roc} fractions $T$, or \protect\subref{fig:d3_cwola_corrected_roc} fractions $R$. Curves for neural-network classifiers are shown as solid lines and curves for traditional substructure observables are dashed lines. In the data-derived panels (\protect\subref{fig:d3_parton_roc}, \protect\subref{fig:d3_cwola_roc}, \protect\subref{fig:d3_cwola_corrected_roc}), TopicFlow generative classifiers assume the same quark fractions used to construct the curves, while in \protect\subref{fig:d3_mc_roc} fractions $S$ are used.}
    \label{fig:rocs}
\end{figure}
\begin{table}
    \centering
        \begin{tabular}{ccccc}
            \toprule
            \multirow{2}{4.2cm}{\vspace{-0.3cm}\bf Classifier/Observable} &
            \multicolumn{4}{c}{{\bf AUC}}
            \\\cmidrule{2-5}
                             & {\bf MC}    & {\bf Data (\emph{S})}  & {\bf Data (\emph{T})} & {\bf Data (\emph{R})} \\\midrule
            Fully Supervised & {\bf 0.768} & 0.658                  & 0.683                        & 0.652               \\
            MC CWoLa         & 0.750       & 0.696                  & 0.728                        & 0.689               \\
            Data CWoLa       & 0.707       & {\bf 0.730}            & {\bf 0.754}                  & {\bf 0.722}         \\
            TopicFlow        & 0.694       & 0.704                  & 0.726                        & 0.704               \\
            $\sigma_1$       & 0.737       & 0.683                  & 0.713                        & 0.677               \\
            $p_T^D$          & 0.738       & 0.629                  & 0.650                        & 0.625               \\
            Multiplicity     & 0.728       & 0.553                  & 0.562                        & 0.551               \\
            Mass             & 0.679       & 0.507                  & 0.508                        & 0.507               \\
            \bottomrule
        \end{tabular}
        
    \caption{AUC scores for classifiers and observables under various constructions of the ROC. Data results $S$, $T$ and $R$ differ by the assumed $f_\mathrm{DoubleMu}, f_\mathrm{Jet}$, per Table.\,\ref{tab:fraction-summary}. The best score in each case is indicated in bold.}
    \label{tab:aucs}
\end{table}

In this section we present classification results in the form of significance improvement characteristic (SIC) curves and area under the ROC (AUC) scores. Like the ROC curve, a SIC curve is constructed from quark and gluon efficiencies across all cuts on the classifier output. For an MC testing sample, the efficiencies (and thus the SIC curve) can be evaluated using parton-level truth labels assigned by the matching algorithm. Of course ultimately we want to rank classifiers by their potential performance on data, not MC. Since truth labels are not present for data, the efficiencies cannot be determined directly. However, one can instead use an estimate of the quark fractions $f_i$ of two mixtures $M_i$ as well as Eq.\,\ref{eq:demix-distributions} to write:

\begin{equation}
    \label{eq:data-roc}
    \begin{gathered}
        \varepsilon_{Q} = \frac{(1-f_2)\varepsilon_{M_1} - (1-f_1)\varepsilon_{M_2}}{f_1-f_2},\\
        \varepsilon_{G} = \frac{f_1\varepsilon_{M_2} - f_2\varepsilon_{M_1}}{f_1-f_2}.
    \end{gathered}
\end{equation}
Again, this relies on sample independence between the two mixtures. If there are non-trivial differences between MC and real data, then the two measures of the efficiencies may disagree. This is particularly relevant for neural-network-based classifiers which may learn MC-specific features during supervised training.


  
In Fig.~\ref{fig:rocs}, we present both MC truth-derived (\ref{fig:d3_mc_roc}) and data-derived (\ref{fig:d3_parton_roc}, \ref{fig:d3_cwola_roc}, \ref{fig:d3_cwola_corrected_roc}) SIC curves for each classifier, shown as solid lines, and the substructure observables discussed in Secion~\ref{sec:datasets}, shown as dashed lines. Here, the label ``TopicFlow" refers to a generative classifier of the form $p_Q(x)/p_G(x)$.
The three data-derived panels assume different $f_\mathrm{DoubleMu}, f_\mathrm{Jet}$ ($S$, $T$, $R$) as given in Tab.~\ref{tab:fraction-summary} . For the neural-network classifiers, curves are produced by averaging predictions of 10 independent models on the test set of the relevant dataset. Curves for the substructure observables are measured over the full dataset. The corresponding AUC scores are summarised in Tab.~\ref{tab:aucs}.

\begin{figure}
    \centering
     \subfloat[\label{fig:d3_topicflow_parton_roc}]{
        \includegraphics[width=0.33\textwidth]{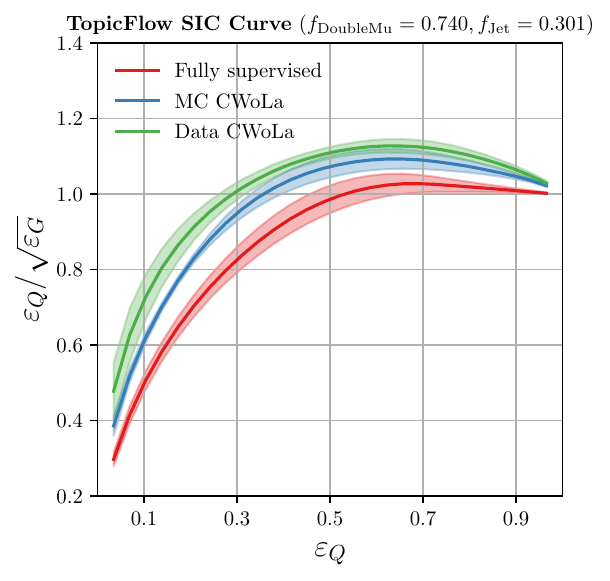}
    }
    \subfloat[\label{fig:d3_topicflow_cwola_roc}]{
        \includegraphics[width=0.33\textwidth]{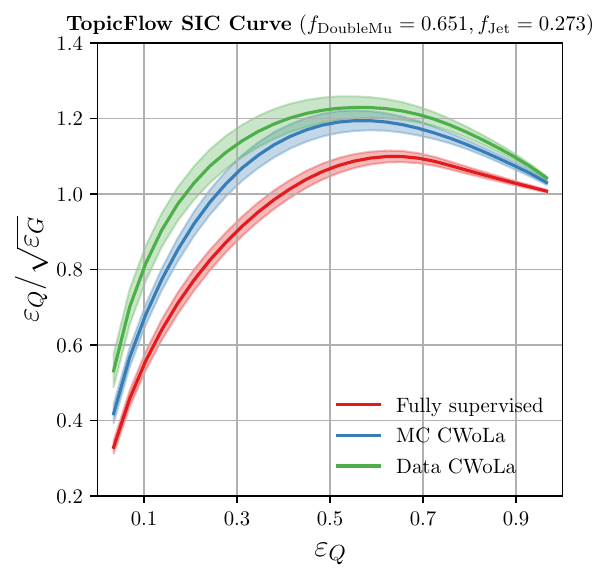}
    }
    \subfloat[\label{fig:d3_topicflow_cwola_corrected_roc}]{
        \includegraphics[width=0.33\textwidth]{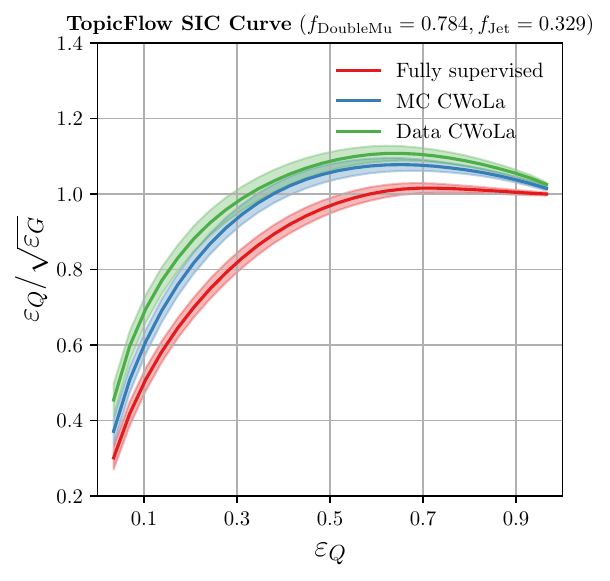}
    }
    \caption{SIC curves constructed using samples from TopicFlow quark and gluon models. Each panel differs by the assumed quark fractions $f_\mathrm{DoubleMu}, f_\mathrm{Jet}$ used to train the TopicFlow model, either \protect\subref{fig:d3_topicflow_parton_roc} $S$, 
\protect\subref{fig:d3_topicflow_cwola_roc} $T$ or
\protect\subref{fig:d3_topicflow_cwola_corrected_roc} $R$
. Error bands show the standard deviation across 100 (10$\times$10) TopicFlow/classifier pairs.}
    \label{fig:topicflow-rocs}
\end{figure}

When evaluating classifiers on MC labels, as for Fig.~\ref{fig:d3_mc_roc}, the fully-supervised classifier is best, closely followed by the MC-CWoLa classifier. Save for jet mass, all of the substructure observables outperform the classifiers trained on data (Data CWoLa and TopicFlow). Considering next the data SIC curves of Figs~\ref{fig:d3_parton_roc}, \ref{fig:d3_cwola_roc} and \ref{fig:d3_cwola_corrected_roc}, the rankings are different. In particular, the Data CWoLa classifier achieves the maximum significance improvement while the fully-supervised network is bested by the $\sigma_1$ observable.\footnote{Note that the peaks at low $\varepsilon_Q$ are likely artefacts due to low statistics in the testing set.} TopicFlow and MC CWoLa achieve similar results to one another and have the next best performance after Data CWoLa. Comparing these data-derived panels to one another, we see that the assumed quark fractions $f_\mathrm{DoubleMu}$ and $f_\mathrm{Jet}$ do not impact the rankings among the classifiers. What does change between these two constructions is the absolute significance improvement of the models as well as the signal efficiency at which it is achieved. Specifically, the maximum SIC is highest in Fig.~\ref{fig:d3_cwola_roc} which uses fractions that assume quarks and gluons are mutually irreducible in Data CWoLa predictions ($T$). On the other hand, the SIC curves in Fig.~\ref{fig:d3_cwola_roc}, which use fractions extracted by Data CWoLa assuming the same $\kappa_{QG}$ as MC, have the lowest maxima. In fact, both panels in the right column indicate that little-to-no significance improvement can be yielded by any cut on the fully-supervised networks. Notably, no classifier in any of the data-derived panels reaches the maximum SIC achieved by the fully-supervised classifier in MC (Fig.~\ref{fig:d3_mc_roc}).

Finally, Fig.~\ref{fig:topicflow-rocs} shows SIC curves of each discriminative classifier as estimated using samples from TopicFlow quark and gluon distributions.\footnote{TopicFlow generative classifiers are excluded since they represent the precise likelihood ratio for their samples and are therefore biased.} We show three panels corresponding to choices $S$, $T$ or $R$ for the fractions $f_\mathrm{DoubleMu}$ and $f_\mathrm{Jet}$. The error bands show one standard deviation across the collection of curves produced by evaluating 10 classifiers on samples from 10 TopicFlow models. These bands therefore capture uncertainty in the TopicFlow model, which we find comprises most of the error. The general shapes of the curves agree with Fig.~\ref{fig:rocs} except at low $\varepsilon_Q$ where the SIC for Data CWoLa previously exhibited large statistical variations. When amplifying the statistics with TopicFlow generated samples, the curves for each classifier are smooth and clearly exhibit a maximum. The TopicFlow SIC curves also maintain the same hierarchy among the three classifiers as Fig.~\ref{fig:rocs}. We note that the width of the bands on each curve are smaller than the differences between panels in Figs.~\ref{fig:d3_topicflow_cwola_roc} and \ref{fig:d3_topicflow_cwola_corrected_roc}. Thus, the largest source of uncertainty in the expected discrimination power for a given quark gluon tagger in our datasets comes from the definition of the classes themselves. Again, as one expects, the greatest peak SIC is predicted by assuming mutually irreducible quarks and gluons since these are by definition maximally-separable categories.

\section{Conclusions}
\label{sec:conclusion}

Machine-learning methods that operate directly on real data offer a way to avoid test-time performance loss caused by imprecision of Monte Carlo simulation. In this work we have studied the application of weakly-supervised neural networks to quark/gluon classification using jets in the CMS Open Data. In particular, we took \zjet and dijet events as respective quark- and gluon- enriched samples to train CWoLa and TopicFlow models. The performance of these models was compared to fully-supervised networks trained on MC simulation.

In addition to the predictions based on parton labels in simulation, we made two estimates for the quark fractions of each real dataset. The first assumes that quarks and gluons are irreducible in the output of the classifiers, in line with the jet-topic framework. The second estimate assumes non-zero reducibility between quarks and gluons under the classifiers, as determined by MC. The results of both estimates differ from the MC prediction. These discrepancies may be covered by systematics that we cannot quantify in the CMS Open Data.

Using the extracted fractions, TopicFlow networks were trained to model the underlying quark and gluon distributions in real data. We employed a variant of the loss function introduced in Ref.~\cite{Dolan:2022ikg} that allows for more stable training due to a convex optimisation objective. The distributions encoded by these TopicFlow models fit the corresponding topics in data better than predictions from simulation---both pure quark/gluon categories and topics constructed in MC.

We constructed SIC curves for a number of classifiers both in MC and in real data, using each of the different estimates of the mixture quark fractions for the latter. We find the rankings of the fully- and weakly- supervised classifiers are reversed when evaluating on data versus MC, irrespective of the choice of fractions. Specifically, CWoLa models trained on real jets achieved the best SIC on data whereas fully-supervised networks were only performant when testing on MC. This highlights the importance of weakly-supervised learning in closing the domain shift between data and simulation. The generative classifiers constructed with TopicFlow performed competitively with CWoLa. We also used TopicFlow to smooth statistical fluctuations in the SIC curves and found that the largest uncertainty on the discrimination power of the models comes from the definition of the categories via the assumed fractions.

An interesting direction for future work would be to relax the assumption of sample independence of quarks and gluons in the mixed datasets. While we observed that sample independence is approximately satisfied in the Monte Carlo, we did not attempt to quantify the effects of its violation. As such, our results are subject to this extra degree of uncertainty. Some ideas to this end are given in Ref.~\cite{Komiske:2018vkc}. A different approach may be to match the two mixed distributions via reweighting as in Ref.~\cite{ATLAS:2023pdx} (and implicitly in Ref.~\cite{Komiske:2022vxg}). Alternatively, techniques that directly morph between two distributions could be used, such as in Refs.~\cite{Algren:2023qnb, Golling:2023mqx, Diefenbacher:2023flw, Butter:2023ira}. Beyond cases where sample dependence is only approximately satisfied, a dedicated treatment of this problem could extend the reach of weakly-supervised approaches to domains where the effect is known to be severe~\cite{Lee:2023tfx}.

\appendix

\section{The CMS Open Data}
\label{sec:TheCMSOpenData}
\subsection{Accessing the Open Data}
\label{sec:AccessingtheOpenData}
The data from the 2010--2012 period can be accessed remotely or downloaded in the Analysis Object Data (AOD) format, a ROOT-compatible collection of data processed through the CMS software framework CMSSW~\cite{CMSSW}. The AODs are grouped into \textit{primary datasets} on the basis of High-Level Trigger (HLT) paths that share common physics content. The typical size of an AOD file is a few gigabytes, and some thousands of AOD files constitute a primary dataset, which typically has a total size on the order of terabytes. Since we are interested in sourcing data samples with enriched quark- and gluon-jet content, our study makes use of the \texttt{Jet}~\cite{jet2011A} and \texttt{DoubleMu}~\cite{doublemu2011A} primary datasets, from which we extract gluon-rich dijet events and quark-rich $Z$~+~jets events, respectively.

Our study of the 2011 Open Data takes place within version 5.3.32 of CMSSW, the recommended version compatible with the 2011 data. This software framework is used in data taking, processing, analysis and the generation of simulated samples. It defines the event model used to store the data and contains services and plugins we use in the initial stages of our data processing. The code is open source and available on GitHub~\cite{CMSSW}.

The CMSSW framework is also available through the CernVM file system (CVMFS)~\cite{blomer_jakob_2020_4114078}. The open data release is accompanied by a virtual machine (VM) image, to which CVMFS is mounted, and this provides a simple interface to access the data and conduct analysis. Rather than using the VM, we instead use a light-weight container image executed on a batch computing system. The image contains the base operating system needed to execute CMSSW. This, along with a CVMFS mount, allows us to run the CMS software framework with only the data and some additional resources we discuss below.

Event processing within CMSSW consists in the data in an event being passed sequentially to a number of user-defined or built-in modules that are loaded at runtime. These modules implement algorithms that read and write to a shared data structure representing a collision event. The project's top-level configuration file governs this process. The file defines which modules are loaded, the order in which they act, how the data are output and the values of various configurable parameters. 

Our approach to accessing the data is to download each AOD file on demand through the XRootD protocol~\cite{xrootd}, pass the events through our selection and analysis code within CMSSW, and write out much smaller, custom data files in the HDF5 format for downstream processing. This workflow allows us to work with multiple primary datasets without keeping additional terabytes of information we do not require in our study.

We use a custom \texttt{EDAnalyzer} module to write out the information relevant to our analysis directly. (That is, we do not use the output module defined in the CMSSW configuration file for further processing of the data.) Our data payload includes values of kinematic variables, correction factors, vertex and trigger information, as well as jet-substructure observables. These are discussed in more detail in the following section.

Our experience using the CMS Open Data has greatly benefited from the detailed former studies using the CMS Open Data and the increasing number of example analyses provided by CMS through the \texttt{cms-opendata-validation} repositories~\cite{open-data-validation}. Specifically, we have drawn on example code for the production of flat jet-tuples using the 2011 data~\cite{2011-jet-inclusivecrosssection-ntupleproduction} and the general procedures of Ref.~\cite{Tripathee:2017ybi} and the \texttt{MODProducer}~\cite{mod}, adapted to work with the prescaled dijet triggers of the 2011 Jet Primary Dataset. We point the interested reader to Refs.~\cite{Tripathee:2017ybi, challenges_and_directions,Bellis:2022onc} for a detailed discussion of challenges and advice for using the CMS Open Data.

\subsection{Analysing the Open Data}
\label{sec:AnalysingtheOpenData}

In addition to the AOD files, our analysis code requires access to the CMS condition database~\cite{Guida_2015,conditiondatabase}. This is a store of auxiliary, non-event-related data provided in the form of SQLite files within a local \texttt{/cvmfs/cms-opendata-conddb.cern.ch} directory inside the container image. In the context of our analysis these provide access to trigger information and to the jet energy corrections (JECs): MC-derived correction factors that mitigate detector noise and pile up. The condition database contains information relevant for different data-taking periods, each labelled by a \textit{global tag}. Our analysis uses the global tags \verb|START53_LV6A1::All| for simulation and \verb|FT_53_LV5_AN1::All| for data. A connection is established to the conditions database within our configuration file.

Before processing the data through our \texttt{EDAnalyzer} module, we first define the set of luminosity sections (or luminosity blocks) to be processed. This is done by reading a \texttt{json} file prepared by the CMS data-quality monitoring group, which lists the validated luminosity sections by run number. The data that we use are collected from luminosity sections within the list of validated runs from 2011 data taking between the run numbers 160,431 and 173,692.

The \texttt{EDAnalyzer} module reads the AOD files and outputs the HDF5 data files we use in our subsequent analysis. Its main role is to write out kinematic variables associated with $Z$-candidate muon pairs and jet constituents, for events which are respectively accepted by the specific dimuon and dijet triggers on which our analysis is based. An array of strings labelling the triggers of interest are passed into our \texttt{EDAnalyzer} module from the configuration file, and a list of active triggers is stored at the beginning of each run, over which trigger information remains constant. For each event we access the trigger results and, if the trigger fired, store information such as the total trigger prescale. This is necessary to construct smooth histograms for events accepted by different triggers. In our study, we concentrate on events from one trigger at a time, and so the prescale values play no role in our analysis.

The jet constituents are the objects obtained from the Particle Flow (PF) algorithm~\cite{CMS-PAS-PFT-09-001,CMS-PAS-PFT-10-001,CMS:2017yfk}. In our analysis code, we pull jets from the \texttt{ak5PFJets} collection, corresponding to Fastjet~\cite{Cacciari:2011ma} anti-$k_t$ jets~\cite{Cacciari:2008gp} reconstructed from the PF particles with $R=0.5$. The constituent and jet $p_T$, along with the jet energy fractions (used later to implement our jet-quality requirements), are scaled by the JECs mentioned earlier. These are sourced using the \texttt{ak5PFL1FastL2L3Residual} correction label for data, and \texttt{ak5PFL1FastL2L3} for MC.

Another layer of processing is required to deal with pileup: contamination from additional, non-event vertices. Here, we apply the procedure of Charged Hadron Subtraction (CHS)~\cite{CMS:2014ata}, where tracks not associated with the event vertex (defined below) are pruned from a jet. We draw vertex information from the standard vertex collection provided by CMSSW within our configuration file with the requirements that the minimum number of degrees of freedom in the vertex fit be larger than 4; $|z| \leq 24$~mm, where $z$ is the $z$-coordinate of the point of closest approach of the tracks to the $z$-axis; and $\rho < 2$, where $\rho$ is an estimate of the average amount of transverse energy in the event per unit area coming from soft activity such as pileup. From among these vertices, the \textit{event vertex} is defined as that primary vertex whose tracks have the largest sum of squared $p_{T}$, while all other interaction vertices are classed as pileup vertices. Within the \texttt{EDAnalyzer} we write out the vertex-fit information associated with the charged hadrons and later remove those assigned to pileup vertices. The distribution of the number of offline primary vertices (NPV) looks different in simulation compared to data for the triggers of interest to our study. To correct this, we apply an additional weight to the MC that makes the NPV distributions match.

The remaining PF candidates are used to calculate the jet observables that form the basis of our machine-learning study. (These are introduced in more detail in Sec.~\ref{sec:training}.) These observables are IRC-safe, and therefore an additional cut on the $p_T$ of the jet-constituent objects is unnecessary.

The muon objects that enter our analysis are required to pass the \texttt{GlobalMuonPromptTight} requirements. These include that the track should be identified as a global muon~\cite{CMS:2012nsv,CMS:2018rym}, the associated normalised $\chi^2$ in the global muon fit should be less than 10, and the number of valid muon-detector hits used in the fit should be non-zero.

In order to associate truth-level parton information with the PF candidates in MC, the generator objects must be matched to the PF objects from offline reconstruction (RECO) through MC matching. We follow the implementation of this matching from the CMSSW Physics Analysis Toolkit~\cite{Adam:1196152}. Specifically, we call the \texttt{MCMatcher} \texttt{EDProducer} within our configuration file. This matches jets from the \texttt{ak5PFJets} collection we use in our analysis to the \texttt{genParticles}, the MC-truth particle collection. The matching is performed for objects satisfying $\Delta R < 0.4$ and a relative $p_T$ difference $|p_T^{(2)} - p_T^{(1)}| / |p_T^{(2)}| < 3.0$, where object (2) is the RECO-object and (1) is the generator object. Additionally, we forbid two RECO objects from matching to the same generator-level object, and preference matches by minimising $\Delta R$. We then write the PDG IDs of the matched partons to our HDF5 files for use later.


\section{Definitions of substructure observables}
\label{app:substructure-definitions}

\subsection*{Energy flow polynomials (EFPs)}

An EFP can be identified with a multigraph where each node gives an energy-weighted sum over particles, and each edge denotes an opening angle between two particles. The polynomial corresponding to a particular graph $G$ with $N$ nodes is
\begin{equation}
	\mathrm{EFP}_G = \sum_{i_1=1}^{M} \cdots \sum_{i_N=1}^{M} z_{i_1} \cdots z_{i_N} \prod_{(j,k)\in G} \theta_{i_ji_k}^{\,\beta}\,,
\end{equation}
where $M$ is the number of jet constituents, $z_a=p_{T,a}/p_{T,J}$ is the $p_T$ fraction of constituent $a$ relative to the jet, $\theta_{ab}$ is the separation of $a$ and $b$ in the rapidity-azimuth plane, and $\beta$ is a parameter. EFPs are manifestly IRC safe.

\subsection*{Jet fragmentation distribution $p_T^D$}
The jet fragmentation distribution is defined as
\begin{equation}
    p_T^D = \sqrt{\frac{\sum_i p_{T,i}^2}{\left(\sum_i p_{T,i}\right)^2}}
\end{equation}
where the sum runs over the jet constituents. This variable gives an output between 0 and 1. If there is only a single constituent in the jet then $p_T^D=1$, and in the limit that there are an infinite number of constituents then
$p_T^D=0$. Accordingly, higher values of $p_T^D$ are associated with quark jets
and lower values with gluon jets.

\subsection*{Jet major axis $\sigma_1$}
The major axis of a jet is determined by calculating the eigenvalues of the matrix
\begin{equation}
    \begin{aligned}
        M_{11} &= \sum_i p_{T,i}^2 \Delta\eta_i^2 \, ,\\
        M_{22} &= \sum_i  p_{T,i}^2 \Delta\phi_i^2 \, ,\\
        M_{12}=M_{21} &= -\sum_i p_{T,i}^2 \Delta\eta \Delta\phi \,.
    \end{aligned}    
\end{equation}
The major axis is defined as the larger eigenvalue $\lambda_1$ of $M$ normalised to the sum of the squared transverse momenta of the jet,
\begin{equation}
    \sigma_1 = \left(\frac{\lambda_1}{\sum_i p_{T,i}^2} \right)^{1/2} \, .
\end{equation}
\section{Effects from parton matching}
\label{app:parton-matching}
In this appendix, we study the effect of removing non-parton-matched jets from the simulated samples.
\begin{figure}
    \centering
    \subfloat[\label{fig:allmatch_preds}]{
        \includegraphics[width=0.33\textwidth]{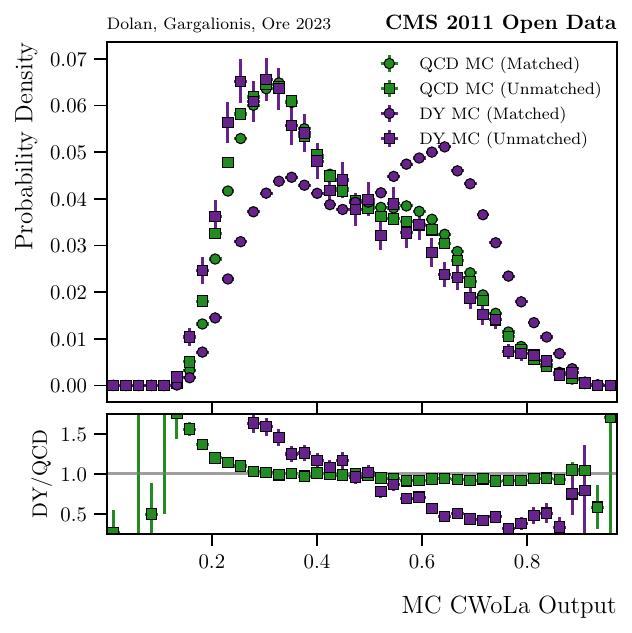}
    }
    \subfloat[\label{fig:allmatch_sigma_1}]{
        \includegraphics[width=0.33\textwidth]{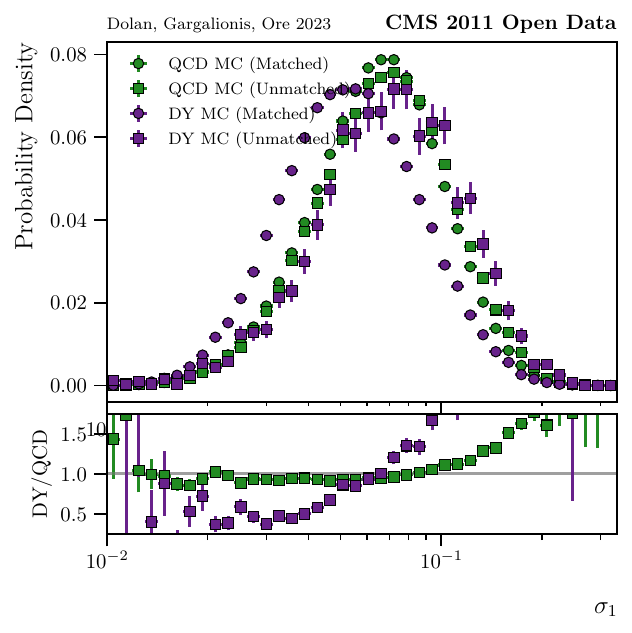}
    }
    \subfloat[\label{fig:allmatch_ptD}]{
        \includegraphics[width=0.33\textwidth]{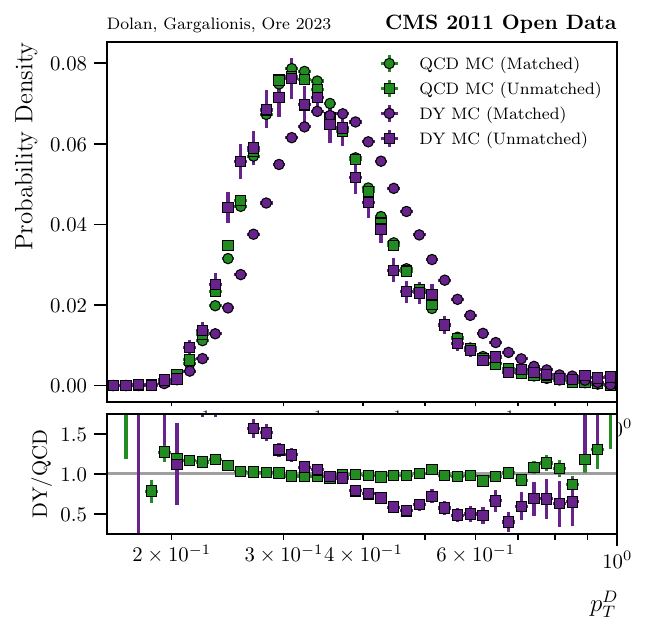}
    }    
    \caption{Histograms of \protect\subref{fig:allmatch_preds} the MC CWoLa network output \protect\subref{fig:allmatch_sigma_1} $\sigma_1$ and  \protect\subref{fig:allmatch_ptD} $p_T^D$ for matched and unmatched jets in QCDSim and DYSim MC.}
    \label{fig:parton-match-validation}
\end{figure}
Figure~\ref{fig:parton-match-validation} shows the distributions of jet substructure observables for the leading jet in QCD and DY MC, split by whether or not the jet has a parton match. We show standard quark/gluon discriminators $\sigma_1$ and $p_T^D$ as well as the MC CWoLa classifier score.

Based on these variables, we identify minimal difference between matched and unmatched jets in QCDSim. On the other hand, we see that the unmatched jets in DYSim behave as jets in QCDSim. This suggests
that the unmatched DYSim sample is enriched in gluon-initiated jets. Consequently, the gluon fraction of the full DYSim sample is likely higher than the estimate based on particle IDs. However, by assuming that the quark fraction of unmatched jets in DYSim is 30\% (i.e. equal to the QCDSim
sample) and taking into account that only 3\% of DYSim jets are unmatched, we expect the true quark fraction in DYSim to be 0.727, just 2\% smaller than our nominal value from Eq.~\ref{eq:mc-fractions}.

We can also check how the classifier performance is affected by training MC CWoLa networks (i.e. DYSim vs QCDSim) including or excluding jets without parton matches. Table~\ref{tab:parton-match-aucs} shows the AUC score for networks trained and evaluated on different subsets of the simulated datasets. For training, we consider either all jets, or only those with a parton match. For testing, we calculate the AUC separately over jets with a match and jets without. In all cases, the AUCs agree within errors, which we estimate as the standard deviation over 30 networks. We therefore conclude that the removal of parton-matched jets from the simulated samples does not have a large effect on the classifier results.

\begin{table}[]
    \centering
    \begin{tabular}{cccc}
    & & \multicolumn{2}{c}{Train set}\\
    \addlinespace[-\aboverulesep] 
     \cmidrule[\heavyrulewidth]{3-4}
    & & Matched  & All \\
    \cmidrule{2-4}
    \multirow{2}{*}{Test set} & Matched  & $0.607 \pm 0.002$ & $0.606 \pm 0.002$ \\
    & Unmatched & $0.607 \pm 0.002$ & $0.605 \pm 0.002$\\
    \cmidrule[\heavyrulewidth]{2-4}
    \addlinespace[-\belowrulesep] 
    \end{tabular}
    \caption{AUC scores for MC CWoLa networks (DYSim vs QCDSim) trained with or without non-parton-matched jets, and evaluated over matched and non-matched jets separately. Errors are the standard deviation over 30 networks.}
    \label{tab:parton-match-aucs}
\end{table}

\section{Comparison with prior work}
\label{app:mit-comparison}

Here we present validation of our workflow by comparing both the data and simulation datasets yielded by our pipeline to those used in Refs.~\cite{Komiske:2019jim}, made accessible through the EnergyFlow package~\cite{energyflow} and various Zenodo records~\cite{komiske2019a,komiske2019b,komiske2019c,komiske2019d,komiske2019e,komiske2019f,komiske2019g,komiske2019h,komiske2019i}.

The authors of Refs.~\cite{Komiske:2019jim} base their analysis on $R=0.5$ anti-$k_t$ jets sourced from the 2011 Jet Primary Dataset and the corresponding QCD simulation samples from CMS. Their focus is on events firing the single-jet \texttt{Jet300} high-level trigger, which is unprescaled. To facilitate a comparison of our processing of the data, we isolate events from this trigger in our raw data -- that is, the data output by our custom \texttt{EDAnalyzer}, discussed above -- and apply a consistent set of baseline selection criteria to both datasets. Specifically, we require jets with $|\eta| < 1.9$ and $p_T \in [375, 1000]~\mathrm{GeV}$, and that pass medium JQC.

\begin{figure}
    \subfloat[]{
        \includegraphics[width=0.4\textwidth]{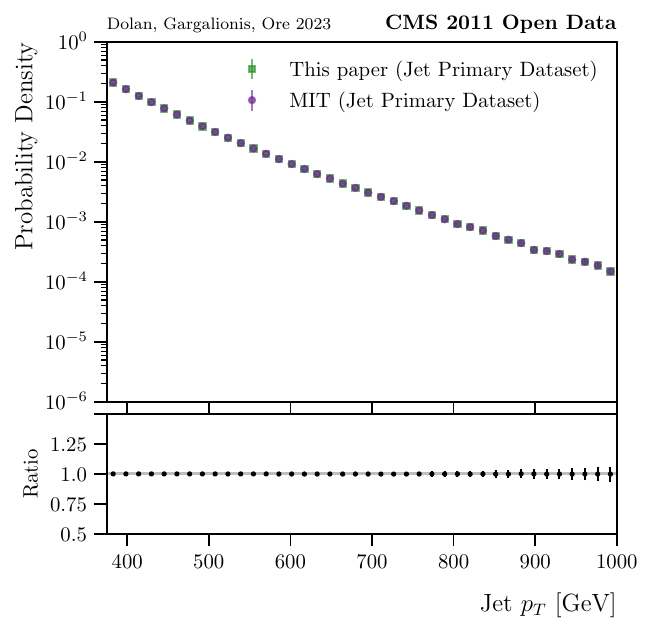}
    }\hspace{1cm}
    \subfloat[]{
        \includegraphics[width=0.4\textwidth]{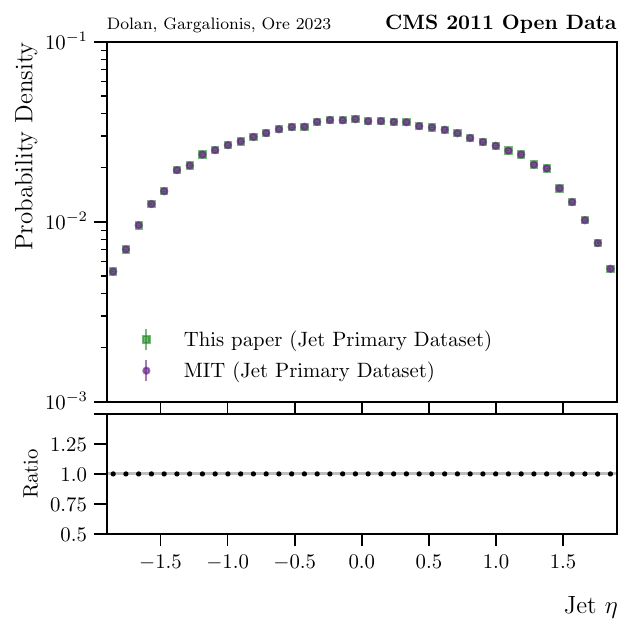}
    }\\
    \subfloat[]{
        \includegraphics[width=0.4\textwidth]{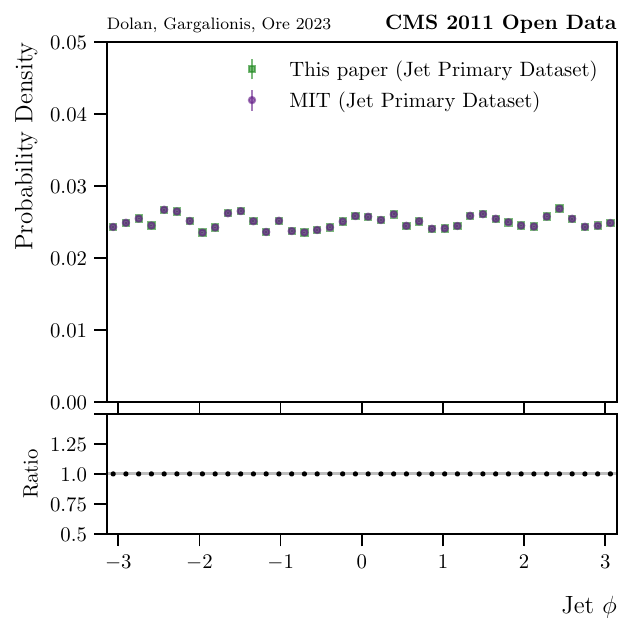}
    }\hspace{1cm}
    \subfloat[]{
        \includegraphics[width=0.4\textwidth]{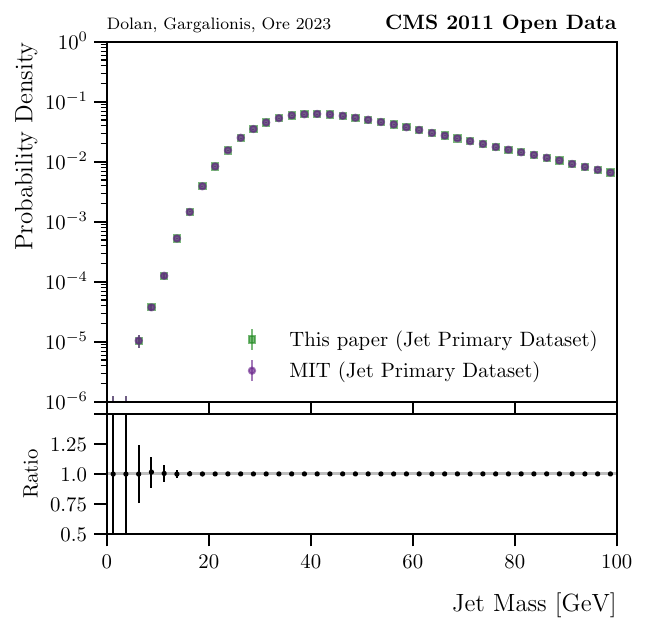}
    }\\
    \subfloat[]{
        \includegraphics[width=0.4\textwidth]{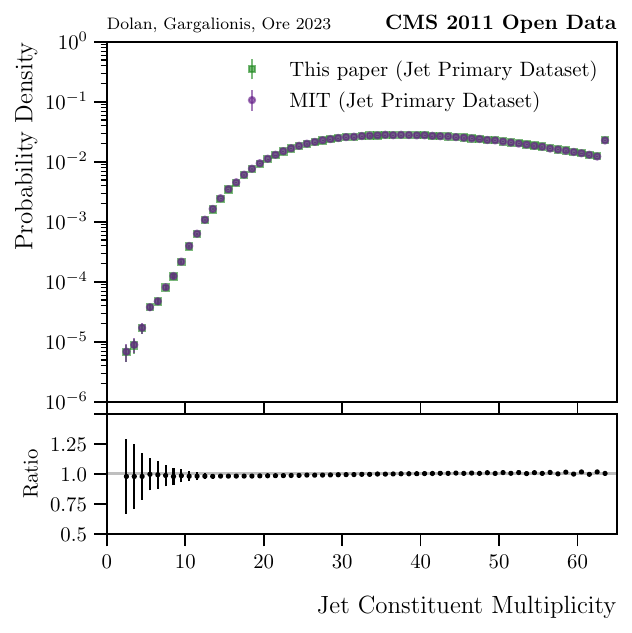}
    }\hspace{1cm}
    \subfloat[]{
        \includegraphics[width=0.4\textwidth]{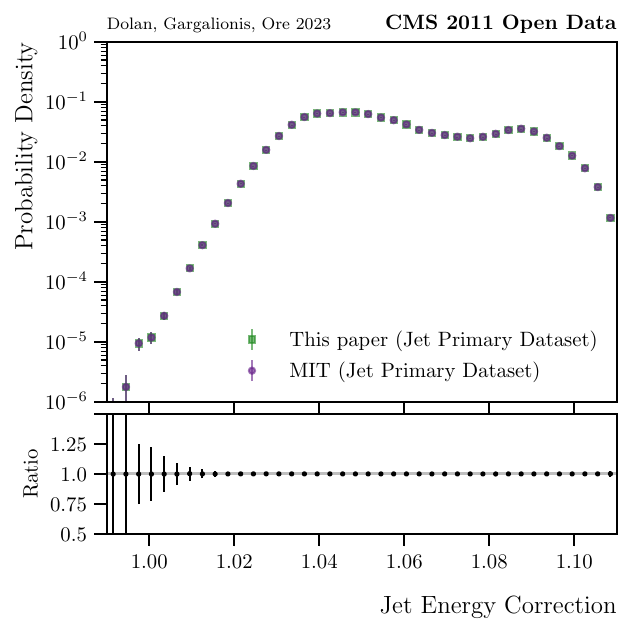}
    }
    \caption{The figure shows a comparison between data sourced from our data pipeline and those from the MIT group~\cite{Komiske:2019jim}.}
    \label{fig:mit_comparison_data}
\end{figure}

\begin{figure}[t]
    \subfloat[\label{fig:mit_comparison_npv_data}]{
        \includegraphics[width=0.4\textwidth]{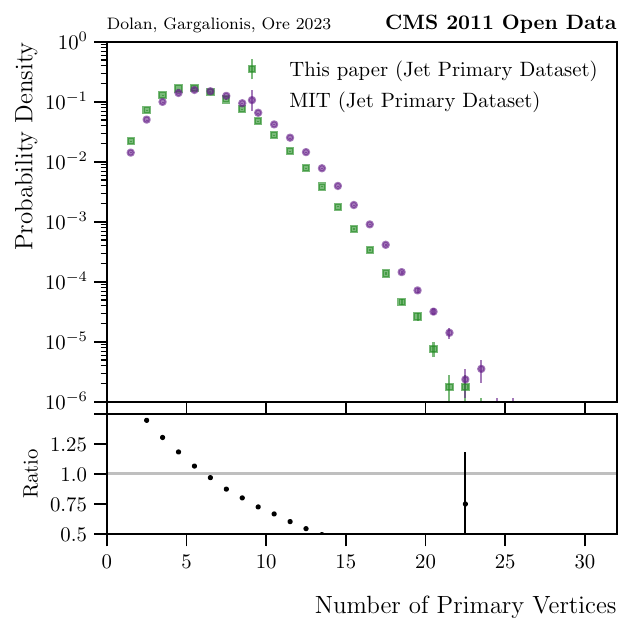}
    }\hspace{1cm}
    \subfloat[\label{fig:mit_comparison_npv_sim}]{
        \includegraphics[width=0.4\textwidth]{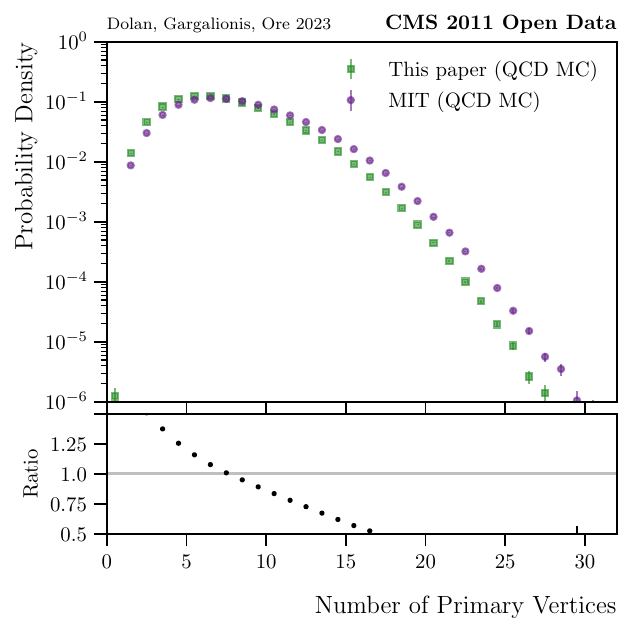}
    }
        
    \caption{The figure shows a comparison of the distribution of the number of primary vertices in (a) real data and (b) simulation between our data pipeline and the data published by the MIT group~\cite{Komiske:2019jim}.}
    \label{fig:mit_comparison_vertices}
\end{figure}

In Fig.~\ref{fig:mit_comparison_data} we show a comparison of jet kinematics, the jet multiplicity and JEC values for data from the CMS Jet Primary Dataset, from both our data pipeline and that of the MIT group. Following Ref.~\cite{Komiske:2019jim}, we keep only the hardest two jets in each event for this comparison. We see that there is good agreement in all of the histograms shown. Additionally, in Fig.~\ref{fig:mit_comparison_npv_data} we show the distribution of the number of primary vertices for the same datasets. Here we note a discrepancy between the distributions, which we understand as coming about from the additional requirements we place on the vertex collection sourced from CMS, described in detail in Sec.~\ref{sec:AnalysingtheOpenData}. For each individual event, our dataset consistently exhibits an equal or lower count of primary vertices compared to the MIT data. The ratio of the NPV distributions in data and simulation is used to mitigate pileup effects through a reweighting of MC histograms. We have checked that the NPV weights derived from the MIT data --- or equivalently, derived from our pipeline without the additional requirements on the vertex collection --- lead to negligible differences in the distributions of kinematic and jet-substructure variables.

\begin{figure}
    \subfloat[]{
        \includegraphics[width=0.4\textwidth]{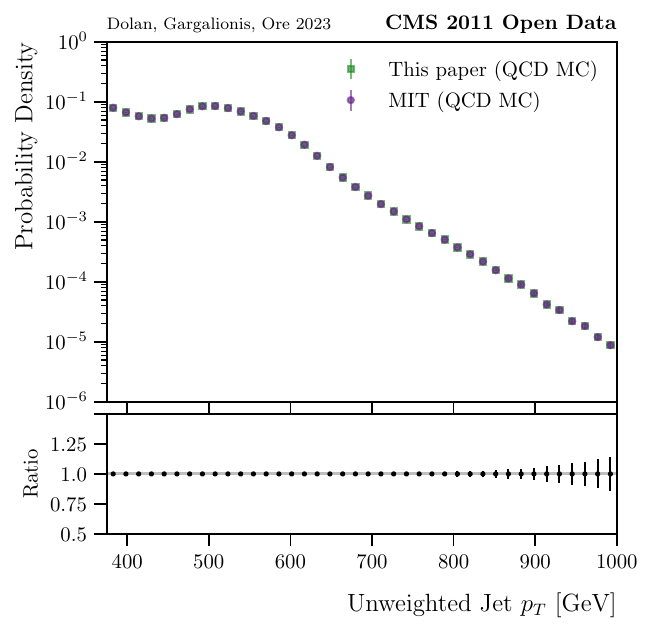}
    }\hspace{1cm}
    \subfloat[]{
        \includegraphics[width=0.4\textwidth]{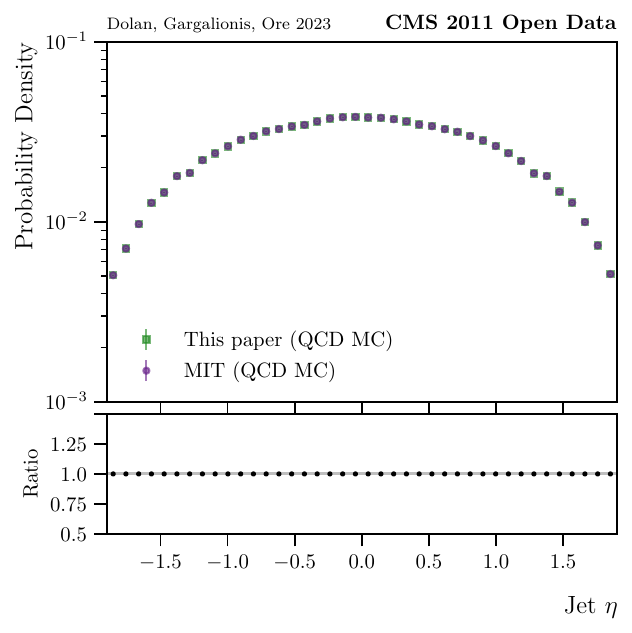}
    }\\[-2.5pt]
    \subfloat[]{
        \includegraphics[width=0.4\textwidth]{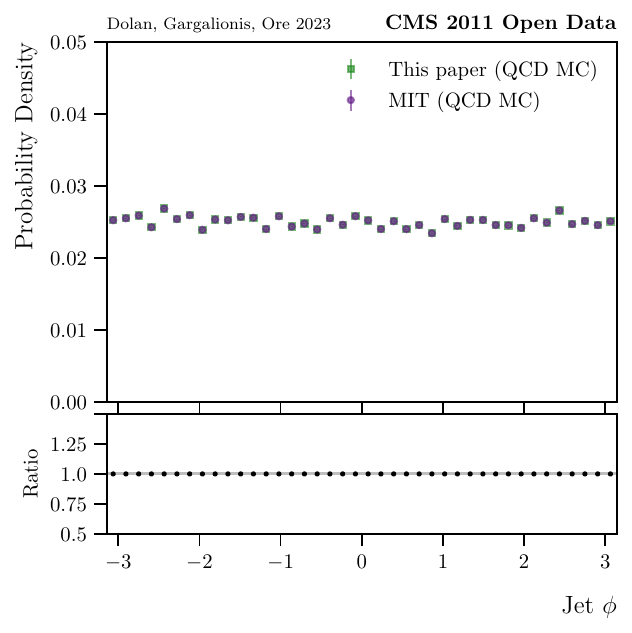}
    }\hspace{1cm}
    \subfloat[]{
        \includegraphics[width=0.4\textwidth]{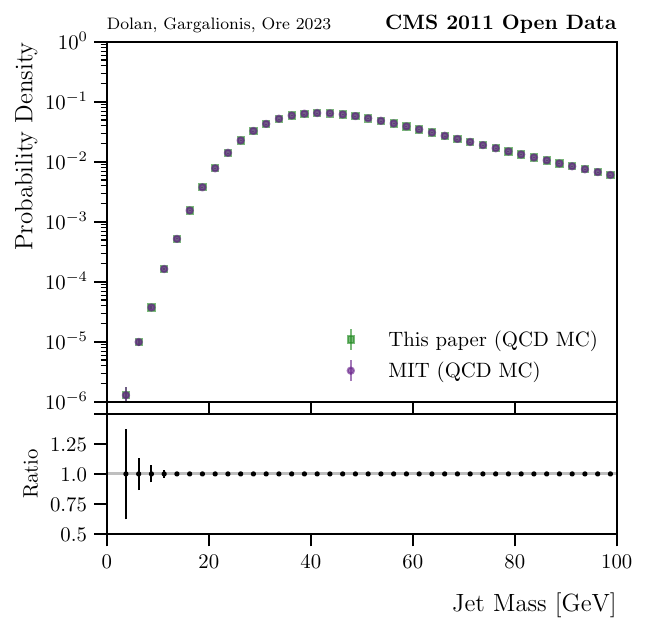}
    }\\[-2.5pt]
    \subfloat[]{
        \includegraphics[width=0.4\textwidth]{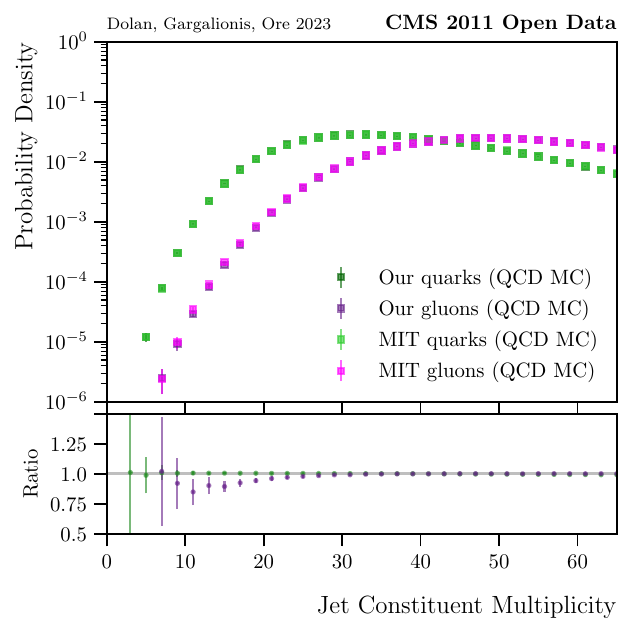}
    }\hspace{1cm}
    \subfloat[]{
        \includegraphics[width=0.4\textwidth]{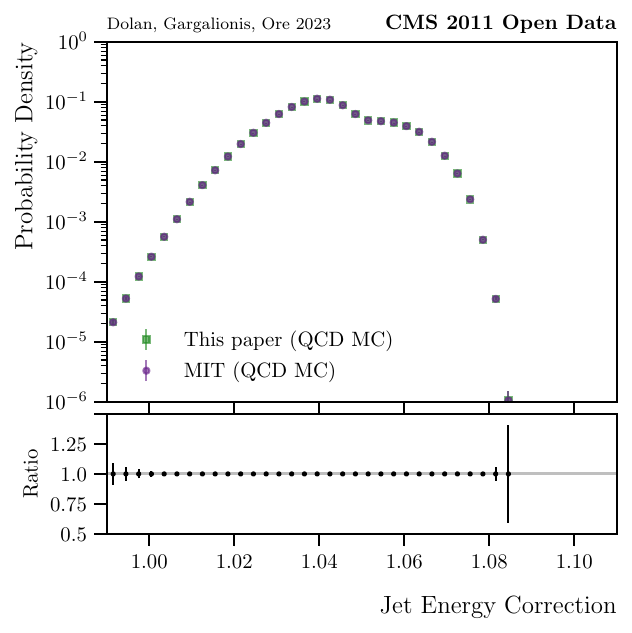}
    }
    \caption{The figure shows a comparison between synthetic data sourced from our data pipeline and those from the MIT group~\cite{Komiske:2019jim}. For the case of the constituent multiplicity we have distinguished between quarks (green) and gluons (purple).}
    \label{fig:mit_comparison_sim}
\end{figure}

In Fig.~\ref{fig:mit_comparison_sim} we show the same comparison for simulated QCD samples from CMS. In this case we include only the hardest jet in each event, sourced from the QCD simulation samples with lower $p_T$ values $300~\mathrm{GeV}$ and $470~\mathrm{GeV}$. All plots are shown without MC weights applied. Again, we see good agreement in all variables shown. In addition, for the comparison in jet constituent multiplicity we distinguish quarks and gluons to show that our methods for parton matching yield similar results. In Fig.~\ref{fig:mit_comparison_npv_sim} we show the distribution of the number of primary vertices in the CMS simulation, which we find differs in a similar way to the data for the same reasons.

\clearpage
\section*{Acknowledgements}

We thank CMS for their commitment to open data and their Open Data team, and in particular Achim Geiser, Julie Hogan, Henning Kirschenmann, Kati Lassila-Perini and Mikko Voutilainen. We also thank Noel Dawe for his collaboration in the early stages of the project. MJD is supported by the Australian Research Council Future Fellowship F180100324. AO is supported by the Deutsche Forschungsgemeinschaft (DFG, German Research Foundation) under grant 396021762 -- TRR 257 \emph{Particle Physics
Phenomenology after the Higgs Discovery}. JG is supported by the ``Juan de la Cierva'' programme with reference FJC2021-048111-I, financed by MCIN\slash AEI\slash 10.13039\slash 501100011033 and the European Union ``NextGenerationEU''\slash PRTR, as well as the ``Generalitat Valenciana'' grants PROMETEO\slash 2021\slash 083 and PROMETEO\slash 2019\slash 087. JG is grateful for the hospitality provided by the Centre of Excellence for Dark Matter Particle Physics at The University of Melbourne during the writing of the manuscript. 

\bibliographystyle{JHEP}

\providecommand{\href}[2]{#2}\begingroup\raggedright\endgroup

\end{document}